\documentclass[a4paper,fleqn,usenatbib]{mnras}

\usepackage[T1]{fontenc}
\usepackage{ae,aecompl}

\usepackage{graphicx,color}   
\usepackage{amsmath}    
\usepackage{amssymb}    
\usepackage{ulem}

\newcommand{\deltad}{\delta_{\rm D}}
\newcommand{\Pnl}{P_\mathrm{nl}}
\newcommand{\Plinj}{P_\mathrm{0}^{(j)}}
\newcommand{\Plinjp}{P_\mathrm{0,+}^{(j)}}
\newcommand{\Plinjm}{P_\mathrm{0,-}^{(j)}}

\title[Power spectrum response of large-scale structure]{Power spectrum response of large-scale structure in 1D and in 3D: tests of prescriptions for post-collapse dynamics}

\author[A. Halle et al.]{Ana\"elle Halle$^{1,2}$,
Takahiro Nishimichi$^{3,4}$,
Atsushi Taruya$^{3,4}$,
St\'ephane Colombi$^{5}$
\newauthor
and
Francis Bernardeau$^{5,6}$
\\
\\
$^{1}$Observatoire de Paris, PSL university, Sorbonne Universit\'e, CNRS, LERMA, F-75014, Paris, France\\
$^{2}$Coll\`ege de France, 11 Place Marcelin Berthelot, 75005 Paris, France\\
$^{3}$Center for Gravitational Physics, Yukawa Institute for Theoretical Physics, Kyoto University, Kyoto 606-8502, Japan\\
$^{4}$Kavli Institute for the Physics and Mathematics of the Universe (WPI), Todai institute for Advanced Study,\\
University of Tokyo, Kashiwa, Chiba 277-8568, Japan\\
$^{5}$Institut d'Astrophysique de Paris, CNRS UMR 7095 and Sorbonne Universit\'e, 98bis bd Arago, F-75014 Paris, France\\
$^{6}$CEA, CNRS, Institut de Physique Th\'eorique, Universit\'e Paris-Saclay, UMR 3681, F-91191 Gif-sur-Yvette, France}

\date{Accepted XXX. Received YYY; in original form ZZZ}

\pubyear{2020}
\begin{document}
\label{firstpage}
\pagerange{\pageref{firstpage}--\pageref{lastpage}}
\maketitle


\begin{abstract}

The power spectrum response function of the large-scale structure of the Universe describes how the evolved power spectrum is modified by a small change in initial power through non-linear mode coupling of gravitational evolution. It was previously found that the response function for the coupling from small to large scales is strongly suppressed in amplitude, especially at late times, compared to predictions from perturbation theory (PT) based on the single-stream approximation. One obvious explanation for this is that PT fails to describe the dynamics beyond shell-crossing. We test this idea by comparing measurements in $N$-body simulations to prescriptions based on PT but augmented with adaptive smoothing to account for the formation of non-linear structures of various sizes in the multi-stream regime. We first start with one-dimensional (1D) cosmology, where the Zel'dovich approximation provides the exact solution in the single stream regime. Similarly to the 3D case, the response function of the large-scale modes exhibits a strong suppression in amplitude at small scales which cannot be explained by the Zel'dovich solution alone. However, by performing adaptive smoothing of initial conditions to identify haloes of different sizes and solving approximately post-collapse dynamics in the 3-streams regime, agreement between theory and simulations drastically improves. We extend our analyses to the 3D case using the PINOCCHIO algorithm, in which similar adaptive smoothing is implemented on the Lagrangian PT fields to identify haloes and is combined with a spherical halo prescription to account for post-collapse dynamics. Again, a suppression is found in the coupling between small- and large-scale modes and the agreement with simulations is improved.

\end{abstract}




\section{Introduction}
\label{sec:Introduction}

A precise quantitative understanding of the Universe is one of the most challenging issues in modern cosmology. In particular, statistical properties of the large-scale matter inhomogeneities are the key to clarify both cosmic expansion history and structure evolution from primordial fluctuations. With upcoming wide-field galaxy surveys such as LSST \citep{LSST2009} or EUCLID \citep{euclid2011}, statistical precision will be greatly improved, and with an accurate theoretical model of large-scale structure described by a set of cosmological parameters, we will be able to tighten the cosmological constraints and to find clues on the nature of dark energy.

Among the techniques to theoretically describe the dynamics and statistics of large-scale structure, cosmological $N$-body simulations allow one to access gravity-induced structure formation in the deeply non-linear regime. However, a large set of simulations is required for an accurate prediction of statistical quantities at large scales, and running $N$-body simulations to explore large parameter spaces remains costly \citep[but see e.g.,][for the so-called emulator approach]{Heitmann_etal2010,Heitmann_etal2009,Lawrence_etal2010,Nishimichi_etal2018_DarkQuest}. In this respect, analytical treatment with perturbation theory (PT) provides a solid framework to efficiently compute statistical quantities, given a set of cosmological parameters. The bottom line of this PT treatment is to solve the evolution of density and velocity fields order by order, based on the single-stream approximation of the Vlasov-Poisson system \citep[see][for a review]{Bernardeau:2001qr}. In this approximation, the cold dark matter distribution is treated as a pressure-less fluid. Strictly speaking, it is valid only during the early phase of structure formation, and is prone to be violated at small scales at later time. Nevertheless, single-stream PT treatments have been shown in practice to accurately describe non-linear mode coupling in the weakly non-linear regime \citep{Jeong:2006xd,Nishimichi:2008ry,Carlson:2009it}, and there have been numerous applications to observations \citep{Blake:2011rj,Oka:2013cba,Beutler17,Zhao19,Ivanov19, dAmico19,Colas20,Troster19}, as well as improved predictions \citep{Crocce:2005xy,Bernardeau:2008fa,Taruya:2007xy,Valageas2007,Matsubara:2008wx,Pietroni:2008jx,Taruya:2010mx}.

It is well-known that the fundamental limitation of single-stream PT appears at the so-called shell-crossing, the collision of matter flows coming from different directions, accompanied by apparent divergences of the density field. Later, the matter flow around shell-crossing regions becomes multi-valued, finally ending up with the formation of virialized structures such as dark matter haloes. Thus, one may expect that single-stream PT ceases to be reliable at scales comparable or below halo sizes. In fact, a direct calculation of higher-order PT corrections suggests a very large ultraviolet (UV) contribution to large-scale modes through non-linear mode coupling, contrarily to N-body results, which means that the break-down of single-stream PT manifests itself even at scales where linear theory predictions are usually trusted \citep{Blas:2013aba,Bernardeau:2012ux}.

So as to better understand the behaviour of PT predictions, \citet{Nishimichi:2014rra} introduced the power spectrum response function, which describes how the power spectrum of large-scale structure of the Universe responds to a small change in initial conditions \citep[see also][for a similar function introduced in the context of local transformations of the density field]{Neyrink_Yang2013}. To be more precise, it is defined as the linear response of the {\it non-linear} power spectrum at wave mode $k$ with respect to the linear counterpart at wave mode $p$, expressed as $K(k,p)$ [see Eq.~(\ref{eq:response_def})]. \citet{Nishimichi:2014rra} found that the response functions measured in $N$-body simulations exhibit a negative amplitude at $k<p$, and that the absolute value of their amplitude is even smaller than that of the single-stream PT predictions if the mode $p$ enters the non-linear regime, indicating a significant suppression of the mode coupling between small and large scales. In other words, the power spectrum in $N$-body simulations is insensitive to the details of the small-scale physics, whereas the single-stream PT predictions generically show UV-sensitive behaviours.

A more precise measurement of the power spectrum response function has then been presented based on a large number of simulations \citep{Nishimichi_etal2017}, quantitatively confirming that a phenomenological damping function needs to be introduced in the single-stream PT prediction in order to account for the suppressed UV sensitivity. While semi-analytic treatment of the response function is proven to be useful to reconstruct the non-linear power spectrum,\footnote{The Python code to reconstruct the non-linear power spectrum, called \texttt{RESPRESSO}, is publicly available at \url{http://www2.yukawa.kyoto-u.ac.jp/~takahiro.nishimichi/public_codes/respresso/index.html}.} the physical origin of the suppressed UV sensitivity in connection with shell-crossing and multi-stream flows still remains unclear. 
The empirical damping factor introduced in \citet{Nishimichi:2014rra} to suppress the strong UV sensitivity of the PT predictions has scale and time dependence given by the condition $\sigma(R;z)=1.35$, where $\sigma$ is the rms dispersion of the linear density contrast smoothed at scale $R$ with a gaussian kernel. This suggests that the breakdown of PT has a connection to the formation of collapsed objects: in the simplest case of spherical-collapse dynamics, haloes form at locations where $\sigma$ reaches $\sim1.69$. However, the connection between the Fourier-space argument of the damped coupling and the configuration-space phenomena of collapsed objects is not trivial.
To clarify this issue, one possible approach consists in performing some (semi-)analytic treatment beyond shell-crossing and to compare the predicted response function with the measured one. In 3D, this is highly non-trivial, partly because even shell-crossing itself is hard to describe with a perturbative treatment \citep[but see][]{Saga_etal2018}, not to mention the subsequent complex evolution of the system. However, one can resort to approximate methods combining ellipsoid collapse dynamics with adaptive smoothing to identify haloes of different masses expected to form at the redshift of interest. This was first proposed by \citet{1996ApJS..103....1B} and exploited later in the public code \texttt{PINOCCHIO} \citep{2002MNRAS.331..587M} which we use in the second part of this article. Using Lagrangian PT to compute displacement fields and a procedure to ``draw'' haloes with some prescribed universal profile depending on their mass \citep[for instance, the so-called NFW profile, e.g.][]{NFW}, \texttt{PINOCCHIO} provides a recipe to account in a simple way for multi-streaming dynamics as a correction to PT.

While the effects of multi-streaming on 3D PT predictions have so far only been approached approximately, the 1D case, discussed in the first part of this article, is particularly enlightening because it can be treated more accurately. Indeed, in 1D, the Zel'dovich approximation \citep[][]{Novikov1969JETP,1970A&A.....5...84Z}, which corresponds to first-order Lagrangian perturbation theory, already provides the exact single-stream solution until shell-crossing. Combining this approach with $N$-body simulations allows us to directly access the origin and impact of shell-crossing on the suppressed UV sensitivity. Furthermore, an analytic description beyond shell-crossing has been recently invented \citep{Colombi:2014lda, Taruya_Colombi2017}. It perturbatively solves the dynamics of multi-stream flows by correcting, at leading order in time, the Zel'dovich solution just after collapse. This post-collapse PT treatment has been explicitly demonstrated to work well until the next shell-crossing time. On top of this, an improved treatment of collapsing haloes employing an adaptive smoothing technique has been proposed \citep{Taruya_Colombi2017}. With this regularization scheme, post-collapse PT (or Zel'dovich approximation, with less good results) was shown to capture well the phase-space structure of haloes in a coarse-grained manner, and this can lead to an accurate prediction of statistical quantities, such as the power spectrum, even in the non-linear regime \citep{Taruya_Colombi2017}.

In this paper, we study the power-spectrum response function both in 1D and 3D. We compare its measurements in $N$-body simulations to semi-analytic prescriptions combining Lagrangian PT with adaptive smoothing procedures to account for post-collapse dynamics. In 1D, an analytic expression for the Zel'dovich case is derived and response functions are computed from $N$-body simulations, Zel'dovich and post-collapse PT treatments, with or without adaptive smoothing. We show that, similarly to the 3D case, the exact single-stream prediction given by the 1D Zel'dovich solution exhibits a strong coupling between small- and large-scale modes, which largely differs from the measurements in $N$-body simulations. On the other hand, when the adaptive smoothing technique is applied, both post-collapse PT and Zel'dovich approximation provide a reasonable agreement with the $N$-body measurements. Best results are obtained with post-collapse PT, as expected. In 3D, we perform detailed comparisons between $N$-body simulations, Lagrangian PT predictions up to third order and the results obtained with \texttt{PINOCCHIO}. Again, the strong mode-coupling seen in the single-stream PT prediction is shown to be suppressed when accounting for multi-stream dynamics, even when performed as approximately as in \texttt{PINOCCHIO}, which provides a reasonable agreement with $N$-body measurements.

This paper is organized as follows. In Sec.~\ref{sec:1D_cosmology}, we present our detailed analyses in the 1D cosmological case. Zel'dovich and post-collapse PT solutions along with adaptive smoothing algorithms are briefly reviewed (Secs.~\ref{sec:zel1d} and \ref{sec:PCPT}), and the set-up of our 1D $N$-body simulations is presented (Sec.~\ref{sec:nbody1d}). Then, we focus on the response function of the power spectrum, by presenting analytical results for the Zel'dovich approximation and the procedure used to perform measurements in $N$-body simulations (Sec.~\ref{sec:resp_func}). This is followed in Secs.~\ref{sec:powev} and \ref{sec:respev} by quantitative analyses of the power-spectrum and the response function. We compare the results obtained from different analytic PT treatments with $N$-body measurements, and study how incorporating adaptive smoothing improves the results. In Sec.~\ref{sec:implications}, we turn to the 3D case. After briefly describing the \texttt{PINOCCHIO} algorithm and the numerical set-up (Secs.~\ref{subsec:pino} and \ref{subsec:3dsetup}), in particular the 3D $N$-body simulations used in this work, we discuss measurements of the power-spectrum (Sec.~\ref{subsec:3dpk}) and perform detailed analyses of the response function (Sec.~\ref{subsec:3dresponse}), paying particular attention to the numerical convergence of \texttt{PINOCCHIO} with respect to the mass resolution and the choice of halo profile parameters. Finally, Sec.~\ref{sec:conclusion} is devoted to the summary of our findings and conclusions.

\section{One-dimensional (1D) cosmology}
\label{sec:1D_cosmology}

In this section, we consider the case of 1D cosmology, in which massive infinite planes orthogonal to the $x$-axis interact through the gravitational force, in an expanding universe. In this configuration, the Lagrangian equations of motion of the planes are
\begin{align}
&\frac{{\rm d}x}{{\rm d}t}=\frac{v}{a},
\label{eq:eom0_x}
\\
&\frac{{\rm d}v}{{\rm d}t}+H\,v=-\frac{1}{a}\nabla_x\phi,
\label{eq:eom0_v}
\\
&\nabla^2_x\phi(x)=4\pi\,G\overline{\rho}_{\rm m}\,a^2\,\delta(x),
\label{eq:eom0_delta}
\end{align}
where $x(t)$ and $v(t)$ are respectively the comoving position and peculiar velocity of each plane, $\phi$ is the gravitational potential, $\overline{\rho}_{\rm m}$ the average matter density, $\delta$ the density contrast, $a$ the expansion factor of the Universe and $H=\dot{a}/a$ the Hubble parameter.

\subsection{Zel'dovich solution}
\label{sec:zel1d}
In one-dimension and in the cold case, the Zel'dovich approximation is known to provide the exact solution for the dynamics of mass elements before shell-crossing \citep[][]{Novikov1969JETP,1970A&A.....5...84Z}. It can be explicitly written as
\begin{align}
x(q;t) = q+\psi(q)\,D_+(t),
\qquad
v(q;t) = a(t) \frac{{\rm d}D_+(t)}{{\rm d}t}\,\psi(q),
\label{eq:Zel'dovich_sol}
\end{align}
where $q$ is the (initial) Lagrangian coordinate and the function $D_+$ corresponds to the linear growth factor satisfying the following equation:
\begin{align}
\left[\frac{{\rm d}^2}{{\rm d}t^2}+2H(t)\frac{{\rm d}}{{\rm d}t}-\frac{3}{2}\frac{\Omega_{\rm m} H_0^2}{a^3(t)}\right]\,D_+(t)=0.
\end{align}
The Zel'dovich solution in Eq.~(\ref{eq:Zel'dovich_sol}) contains an arbitrary function $\psi(q)$ which we call displacement field. It is related at very early time $t_{\rm ini} \to 0$ to the linear density field $\delta_{\rm L}(q)$ through
\begin{align}
\frac{{\rm d}\psi(q)}{{\rm d}q}\,D_+(t_{\rm ini})=-\delta_{\rm L}(q;\,t_{\rm ini})=-\delta_{\rm L}(q)\,D_+(t_{\rm ini}).
\label{eq:def_deltaL}
\end{align}
\subsection{Post-collapse PT solution and adaptive smoothing}
\label{sec:PCPT}
Post-collapse PT \citep{Colombi:2014lda,Taruya_Colombi2017} allows one to follow the evolution of the system shortly after shell-crossing by estimating a correction to Zel'dovich motion due to the back-reaction in the multivalued region. Here, without entering into details of the intricate expressions presented in \citet{Taruya_Colombi2017}, we sketch the main concepts of this modelling which is asymptotically exact when approaching collapse time. Suppose that shell-crossing happens at local Lagrangian position $q_0$, corresponding to a local peak in the linear density field, $\delta_{\rm L}(q)$. Just after shell-crossing time $t_0$, a small multi-stream region develops around $q_0$. In this region, which extends over some time-dependent interval $q-q_0 \in [-\hat{q}_{\rm c}(t),\hat{q}_{\rm c}(t)]$ [with $\hat{q}_{\rm c}(t_0)=0$], the flow is symmetric and three-valued. The key point is that the coordinates of the phase-space sheet elements can be locally expanded as third order polynomials of $q-q_0$ (with time-dependent coefficients), allowing one to analytically solve the multivalued problem $x(q;t)=y$, hence to estimate the force field inside (and outside) the multivalued region as a function of time and to correct pure Zel'dovich motion by integrating the corresponding equations of motion. Formally, the post-collapse PT solution can be expressed as
\begin{align}
x(q;t) &= x_{\rm\small ZA}[q;\hat{t}_{\rm c}(q)]+ \Delta x[q;t,\hat{t}_{\rm c}(q)],
\nonumber
\\
v(q;t) &= v_{\rm\small ZA}[q;\hat{t}_{\rm c}(q)]+\Delta v[q;t,\hat{t}_{\rm c}(q)],
\label{eq:post-collapse PT_sol}
\end{align}
in the multivalued region, $|q-q_0|\leq \hat{q}_{\rm c}(t)$. In these equations, $\hat{t}_{\rm c}(q) \equiv \hat{q}_{\rm c}^{-1}(q)$ is the inverse of the function $\hat{q}_{\rm c}(t)$: it represents the time when the fluid element with initial position $q$ enters the multivalued region. Functions $x_{\rm\small ZA}$ and $v_{\rm\small ZA}$ are the Zel'dovich solutions given by Eq.~(\ref{eq:Zel'dovich_sol}). Functions $\Delta x[q;t,\hat{t}_{\rm c}(q)]$ and $\Delta v[q;t,\hat{t}_{\rm c}(q)]$ include, in addition to the Zel'dovich displacement from time $\hat{t}_{\rm c}(q)$ of the centre of the multivalued region, the internal motion induced by the force field derived from the three-valued flow. This contribution is mainly described by a perturbative polynomial form of the Lagrangian position $q-q_0$.\footnote{in addition, in the outer part of the multivalued region, a term proportional to $[\hat{q}_{\rm c}-(q-q_0)^2]^{(i+3)/2}$ contributes, with $i=2,0$ for the position and the velocity, respectively.} We refer to Sec.~3.3 of \citet{Taruya_Colombi2017} for detailed expressions, which depend only on the local structure of the local density peak: its position, height and second derivative.

Rigorously speaking, the corrections brought by post-collapse PT are only asymptotically exact when approaching collapse time, but practical measurements show that they also provide a rather accurate description of the dynamics of the inner part of the multivalued region up to next crossing time. While it would be possible in principle to proceed iteratively to follow the evolution of the system during successive dynamical times, as proposed in \citet{Colombi:2014lda} for the non-cosmological case, post-collapse PT is still not able to account for mergers. However, \citet{Taruya_Colombi2017} proposed an algorithm based on an adaptive smoothing procedure to describe, at the coarse level, the population of haloes formed at a given redshift. The idea is to summarize a complex halo resulting from multiple mergers with a ``S'' shape structure in phase-space matching, at the coarse level, the intricate structure of the halo. This implies position dependent coarse-graining to locally account for various states of non-linear evolution.

To be more specific, the procedure proposed in \citet{Taruya_Colombi2017} can be described in 3 steps:
\begin{enumerate}
\item[(i)] {\em Smoothing with different cut-offs:} the first step consists in smoothing the initial density at various scales by employing a sharp-$k$ filter function in Fourier space with a varying cut-off wavenumber $k_{\rm cut}$, to obtain a smooth density $\delta_{{\rm L},k_{\rm cut}}$.
\item[(ii)] {\em Dynamical evolution and identification of haloes:} the calculation of the Zel'dovich solution is performed for each field $\delta_{{\rm L},k_{\rm cut}}$. This allows one to identify, for each value of $k_{\rm cut}$, critical points $q_0$ where haloes are susceptible to form and the extension $\hat{q}_{\rm c}$ of the regions which they cover in Lagrangian space. Because we noticed that post-collapse PT performs well until next crossing-time $t_{\rm next}$, haloes of interest are those susceptible to reach $t_{\rm next}$ at the time $t$ of interest, i.e. we require $t \geq t_{\rm next}$. For each of these haloes, we compute post-collapse dynamics in the Lagrangian interval $q-q_0 \in [-\hat{q}_{\rm c}(t),\hat{q}_{\rm c}(t)]$ and tag the halo.
\item[(iii)] {\em Mergers:} to account for mergers and resolve the so-called cloud-in-cloud problem, step (ii) is performed from the largest to the smallest scale, that is for increasing values of $k_{\rm cut}$. At a given step, we account for haloes identified with the procedure described in (ii) only if their centre does not fall in a region already tagged by a halo corresponding to a smaller value of $k_{\rm cut}$. In practice this enforces $t \simeq t_{\rm next}$ in (ii) instead of $t \geq t_{\rm next}$, since we consider all the possible integer values of $k_{\rm cut}$ in the procedure (which makes it costly). At the last step, corresponding to the largest value of $k_{\rm cut}$, all the untagged collapsed structures are accounted for and followed with post-collapse PT, that is the condition for selecting a halo is $t \geq t_0 $ and not $t \geq t_{\rm next}$, in addition to its centre not being already tagged. All the remaining untagged regions are followed with Zel'dovich dynamics at the finest level, since it is exact before shell-crossing.
\end{enumerate}
The procedure above can just be applied to the Zel'dovich fields themselves without accounting for post-collapse dynamics corrections in the intervals $[-\hat{q}_{\rm c}(t),\hat{q}_{\rm c}(t)]$. In this case, because Zel'dovich dynamics fails quicker beyond shell-crossing, it is better to follow the internal evolution of each halo beyond collapse only for half a crossing time instead of a full one in step (ii) above. More specifically, Zel'dovich dynamics with adaptive smoothing was found to give good results when imposing $\tau-\tau_0 \geq (\tau_{\rm next}-\tau_0)/2$ instead of $\tau \geq \tau_{\rm next}$, where $\tau$ is the ``superconformal'' time $\tau\equiv \int {\rm d}t'/a(t')^2$, $\tau_0$ and $\tau_{\rm next}$ its values at shell-crossing and next crossing, respectively.

Note, finally, that after implementing the adaptive smoothing, the predicted mass distribution in phase space is generally discontinuous. This is because we collect perturbative solutions for the displacement field in phase space coming from different coarse-graining scales, without imposing smoothness at the Lagrangian boundaries  between these solutions. To be precise, the discontinuities can appear at the transitions between tails of the ``S'' shape representing a halo and a non-collapsed region, or at the transitions between tails of two ``S'' shapes in the case of a merger. These regions of phase space have a small density contrast and a small spatial extent, and thus weakly contribute to the power spectrum, mostly at scales below the range of our interest. However, the measurements in Sec.~\ref{sec:respev} suggest that they might represent an important source of noise on the response function, which is indeed sensitive to very small changes in fluctuations of the density field.

\subsection{$N$-body simulations}
\label{sec:nbody1d}
In order to numerically resolve cosmological gravitational dynamics in 1D, we use the public $N$-body code \texttt{Vlafroid}\footnote{The \texttt{Vlafroid} code can be found through \url{www.vlasix.org}.} presented in detail in \citet{Taruya_Colombi2017}. This particle-mesh code computes the evolution of $N_{\rm p}$ particles on a periodic grid with a fixed number of cells $N_x$ to solve the Poisson equation by Fast Fourier Transform after calculation of the projected density with Cloud-in-Cell interpolation \citep{1988csup.book.....H}. Time integration is performed using a predictor-corrector scheme with a slowly varying time-step \citep[constraints on this latter are detailed in ][]{Taruya_Colombi2017}. In addition to evolving the particle distribution through Vlasov-Poisson dynamics, \texttt{Vlafroid} can output results obtained with the Zel'dovich approximation and post-collapse PT, with and without adaptive smoothing, as described in the previous section.

Following \citet{McQuinn:2015tva} and \citet{Taruya_Colombi2017}, initial conditions are given by a random Gaussian field with the following power spectrum,
\begin{align}
P_{\rm 1D}(k)=\frac{k^2}{2\pi}\,P_{\rm 3D}(k),
\label{eq:initial_power_spec}
\end{align}
where $P_{\rm 3D}$ is the linear 3D matter power spectrum obtained with the transfer function of \citet{EisensteinHu1998}. A small perturbation is applied to $P_{\rm 1D}(k)$ to compute the response function, as explained in Sec.~\ref{sec:resp}. The cosmological parameters are those of the concordance $\Lambda$CDM model determined by Planck \citep{Planck2015_XIII}. Initial positions and velocities are obtained from the Zel'dovich approximation, at an initial redshift $z=99$, and the simulations are run up to redshift zero. To perform the simulations, we use $N_{\rm p}=163 \, 840$ particles and a spatial resolution corresponding to $N_x=16 \, 384$ in a periodic box of size $L=1 \, 260$~Mpc.
As advocated by \citet{Taruya_Colombi2017}, we perform a high-$k$ cut-off of the initial power-spectrum at integer wavenumber $n_{\rm c}=N_x/10=1 \, 638$, which corresponds in physical units to $k_{\rm c}=(2 \pi/L) n_{\rm c}=8.2$~Mpc$^{-1}$. The factor 10 between particle number and spatial resolution, along with this smoothing of initial conditions, warrants a well defined, smooth evolution of the phase-space distribution function, particularly in the early phases of the history of the system.

\subsection{Response function}
\label{sec:resp_func}

The response function introduced in \citet{Nishimichi:2014rra} quantifies mode-coupling during the non-linear evolution of large-scale structure. To be precise, it describes the linear response, and its amplitude characterizes the strength of the mode coupling between large- and small-scale Fourier modes.

To properly define the response function, we first recall that the non-linear power spectrum can be viewed as the non-linear response of  gravitational evolution to the input power spectrum for a given cosmological model. Mathematically, the non-linear power spectrum at redshift $z$, $P(k;\,z)$, can be expressed as a {\it functional} of the linear power spectrum at the same redshift, $P_0(k;\,z)$, for a given set of cosmological parameters, $\vec{\theta}$.
Then, consider a variation in the input power spectrum around a fiducial cosmological model $\vec{\theta}_{\rm fid}$, which we denote by $\delta P_0(k)$. The non-linear outcome of its variation, $\delta P(k)\equiv P(k)-P_{\rm fid}(k)$ with $P_{\rm fid}$ being the non-linear power spectrum in the fiducial model, can be expressed in general as:
\begin{eqnarray}
\delta P(k) &=& \int \mathrm{d}\ln p \, K^{(1)}(k,p;\vec{\theta}_{\rm fid}) \delta P_0(p)\nonumber\\
  &+&\frac{1}{2}\int \mathrm{d}\ln p_1\int \mathrm{d}\ln p_2 \, K^{(2)}(k,p_1,p_2;\vec{\theta}_{\rm fid}) \nonumber \\
  & & \quad \quad \quad \quad \quad \quad \quad \quad \quad \delta P_0(p_1)\delta P_0(p_2)\nonumber\\
&+&\dots,
\label{eq:pnl_expansion}
\end{eqnarray}
Here, we omit the redshift dependence for simplicity. Note that the variation $\delta P_0$ is not necessarily small. In the above, the kernels, $K^{(n)}$, characterize the non-linear response to the variation imposed in the linear power spectrum under the fiducial cosmology. In this paper, we are particularly interested in the first term, and drop the superscript for simplicity:
\begin{equation}
K(k,\,p; \vec{\theta}_{\rm fid}) = K^{(1)}(k,\,p; \vec{\theta}_{\rm fid}) \equiv p \,\left.\frac{\delta\,P(k;\vec{\theta})}{\delta\,P_0(p;\vec{\theta})}\right|_{\vec{\theta}=\vec{\theta}_{\rm fid}}.
\label{eq:response_def}
\end{equation}
where $\delta$ no longer means variation, but stands for the operation of a functional derivative. We can construct an estimator to measure the response function in numerical experiments based on this definition.

\citet{Nishimichi:2014rra} and \citet{Nishimichi_etal2017} compared the response function predicted by PT at various orders to the measurement in $N$-body simulations in the 3D case. In this section, we compare 1D $N$-body simulation results to Zel'dovich approximation and post-collapse PT, with or without the adaptive smoothing technique. We start by giving the analytical prediction obtained from Zel'dovich dynamics. Then we detail the procedure used to perform measurements of function $K(k,p;z)$ in large sets of simulations.
\subsubsection{Zel'dovich response function}
\label{sec:zelres}
In 1D cosmology, an analytic expression can be obtained for the response function corresponding to the Zel'dovich solution. The Zel'dovich power-spectrum reads \citep[e.g.,][]{BondCouchman1986,1995MNRAS.273..475S,McQuinn:2015tva}:
\begin{align}
P^{\rm (1D)}_{\rm ZA}(k)=\int_{-\infty}^{+\infty} {\rm d}q\,e^{ikq}\,\left[e^{-k^2\,\{I(0)-I(q)\}}-1\right],
\label{eq:pk_ZA}
\end{align}
where the function $I(q)$ is given by
\begin{align}
I(q)&=\int_{-\infty}^{+\infty}\frac{{\rm d}p}{2\pi}e^{-ipq}\frac{P_0(p)}{p^2}
=\int_{0}^{+\infty}\frac{{\rm d}p}{\pi}\cos(pq)\frac{P_0(p)}{p^2}.
\label{eq:func_Iq}
\end{align}
With the definition (\ref{eq:response_def}), using equation (\ref{eq:pk_ZA}) and the symmetry $I(-q)=I(q)$, one can obtain the following expression for the response function in the Zel'dovich approximation:
\begin{align}
K^{\rm (1D)}_{\rm ZA}(k,\,p)&=p\,e^{-k^2I(0)}\,\delta_{\rm D}(k-p) -\frac{1}{\pi}
\frac{k^2}{p} P^{\rm (1D)}_{\rm ZA}(k)
\nonumber\\
&+\frac{1}{2\pi}\frac{k^2}{p} \Bigl[\int_{-\infty}^{+\infty}{\rm d}q\,e^{i(k-p)q}e^{-k^2I(0)}\Bigl\{e^{k^2I(q)}-1\Bigr\}
\nonumber
\\
&+\int_{-\infty}^{+\infty}{\rm d}q\,e^{i(k+p)q}e^{-k^2\,\{I(0)-I(q)\}}\,\Bigr].
\label{eq:kernel_ZA_1Dmain}
\end{align}
Derivation of these analytic expressions is presented in Appendix \ref{app:predictions_zeldovich}.

\subsubsection{Measurement of response functions: procedure}
\label{sec:resp}

As in \citet{Nishimichi:2014rra} and \citet{Nishimichi_etal2017}, we measure the response function in the numerical simulations presented in Sec.~\ref{sec:nbody1d} through the discretised estimator of Eq.~(\ref{eq:response_def}):
\begin{equation}
\hat{K}_{i,j}\Plinj = \frac{\Pnl^{(i)}[\Plinjp]-\Pnl^{(i)}[\Plinjm]}{\left[\left({\Plinjp}-{\Plinjm}\right) / \Plinj \right] \left(\Delta p/p\right)},
\label{eq:response_estimator}
\end{equation}
where the power spectra are averaged on wave-number bins $i$ and $j$, corresponding respectively to the $k$ and $p$ modes of $K(k,\,p;\,z)$. $\Pnl^{(i)}[\Plinjp]$ and $\Pnl^{(i)}[\Plinjm]$ are the non-linear power-spectra in bin $i$ obtained from initial conditions perturbing positively (respectively negatively) the linear power-spectrum in bin $j$ ($\Plinjp$, respectively $\Plinjm$) and $\Plinj= 0.5 \, (\Plinjp + \Plinjm)$. The quantity $\Delta p$ is the width of wave-number bins for $p$ mode.

We study the response function in the interval $[0,1]$~Mpc$^{-1}$, with $k$ and $p$ bins of identical size $\Delta p=0.01$~Mpc$^{-1}$. With our choice of the box size, $L=1260$~Mpc, the cut-off mode $k_{\rm c}$ used to regularize initial conditions is large enough compared to $1$~Mpc$^{-1}$ so that it does not affect the dynamics in the wavenumber interval we consider.
Also, the box size is such that the number of modes per bin $N_{p \,{\rm per \, bin}}= L\, \Delta p / (2 \pi)$ is equal to 2. We sparsely sample the $[0,1]$~Mpc$^{-1}$ interval by choosing 25 $p$ bins (in which the power-spectrum is perturbed) centred at $ \pi/L + (0.5+4n) \Delta p $, with $n$ in ${0, ..., 24}$.

A large number of realizations is required to reduce the noise because, in one dimension, one modulus $k$ corresponds to only two vectors $+{\bf k}$ and $-{\bf k}$. We therefore run a large number of pairs of simulations in which the initial power-spectrum is perturbed of $\pm 3 \%$ in the same $p$ bin, with the same random number generator seed for each pair of perturbed simulations. The average response at wavenumber $k$ to a perturbation at wavenumber $p$ is computed as the average of the response functions for each pair of simulations obtained with the estimator (\ref{eq:response_estimator}). For each of the 25 perturbed $p$ bins, we perform $378\,000$ pairs of simulations with adaptive smoothing for Zel'dovich and post-collapse PT solutions, and $251\,800$ pairs of simulations without adaptive smoothing.

\subsection{Power spectrum in one dimension}
\label{sec:powev}
\begin{figure*}

\resizebox{\hsize}{!}{
\begin{tabular}{c}
\includegraphics{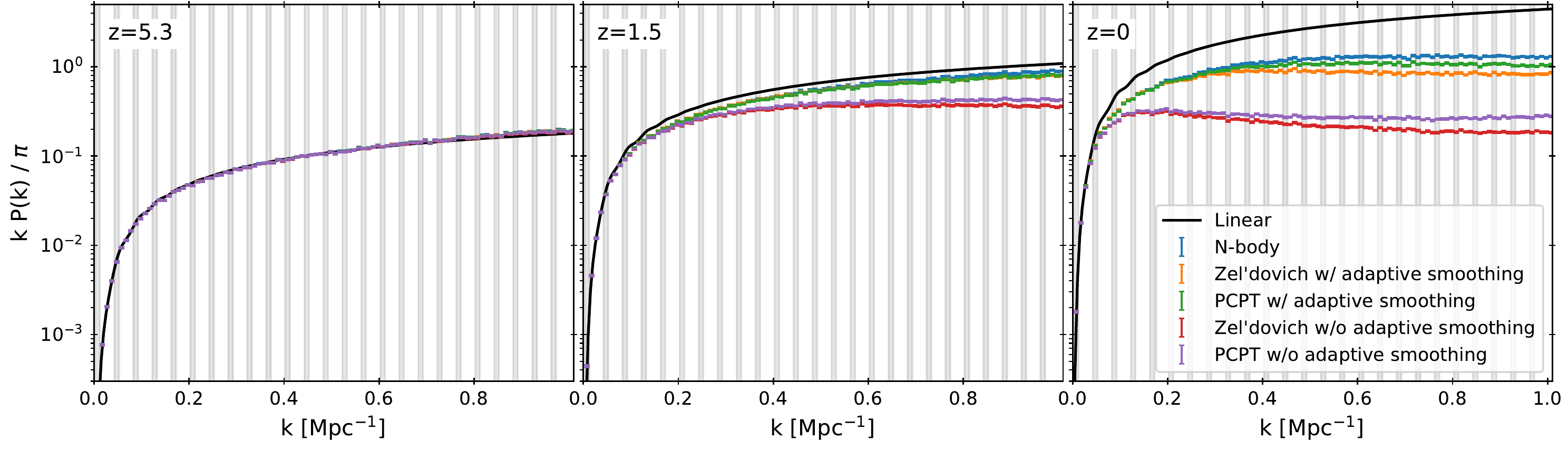} \\
\includegraphics{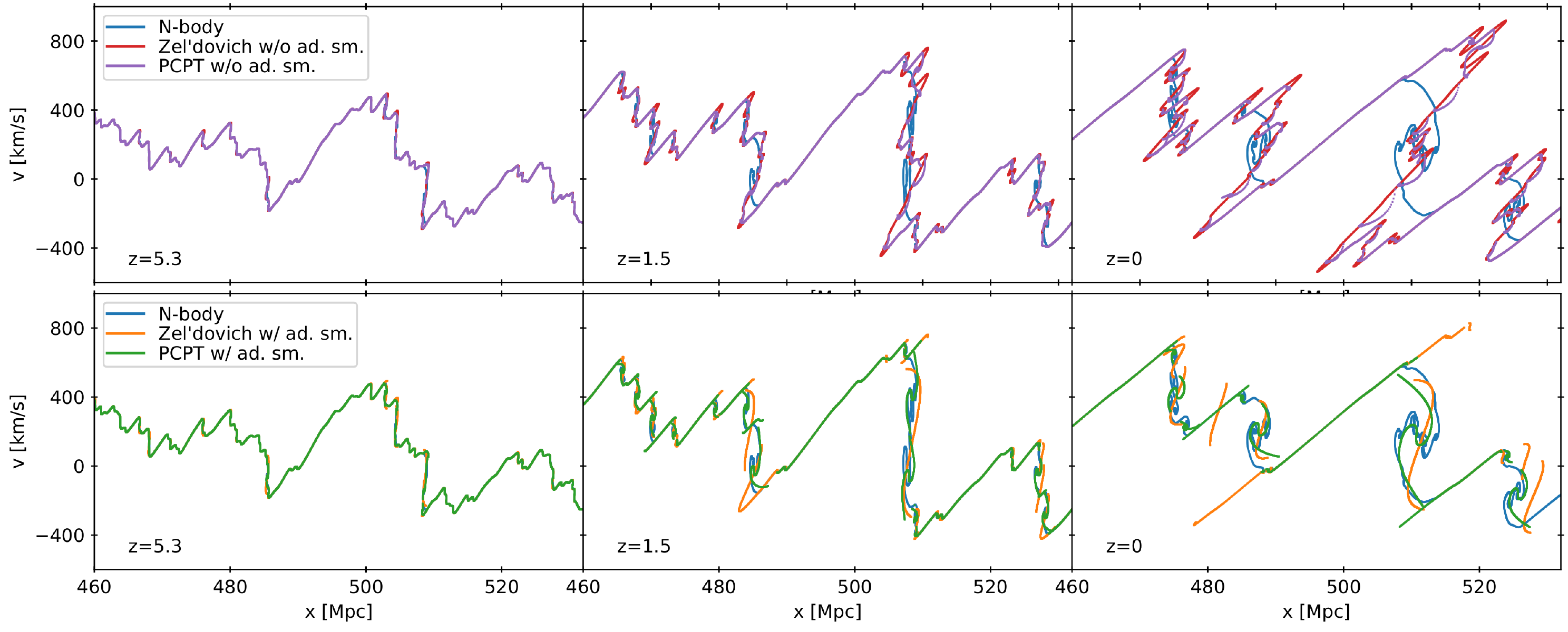} \\
\end{tabular}}

\caption{Top panels: Power spectra obtained at three redshifts from the average over 1000 realizations. The bins used for the computation of the response function are shown in grey. Bottom panels: a region of phase-space in one of the realizations at the same three redshifts. The abbreviations ``w/o'' and ``w/'' stand for ``without'' and ``with'', respectively, ``ad. sm.'' means ``adaptive smoothing'' and PCPT refers to ``post-collapse perturbation theory''.}
\label{pk-fig}
\end{figure*}
Linear, $N$-body, Zel'dovich and post-collapse PT power-spectra are shown at three redshifts on the top panels of Fig.~\ref{pk-fig}, for initial conditions corresponding to unperturbed initial power-spectra. Power-spectra are sampled on bins of the same width $\Delta p = 0.01$~Mpc$^{-1}$ as the ones used for the response function (highlighted as grey bands), and are averaged on a thousand realizations with different random number generator seeds.

At redshift $z=5.3$, the power spectra are all very close to the linear theory prediction. Non-linear effects become significant at $z=1.5$ and differ from the 3D case. In 1D, the amplitude of the power spectrum is damped instead of being enhanced, which may have non-trivial consequences on the properties of the response function, that we study in next section, especially at large $k$'s (response of the small scales) and large $p$'s (response to the small scales). The power-spectrum follows some stable clustering properties at large $k$ \citep{Joyce2011,Benhaiem2013,Taruya_Colombi2017}, visible as a plateau of $k \, P(k)$ for $k \ga 0.5$ Mpc$^{-1}$ and $z=0$. Note that what we call here ``stable clustering'' is very specific in the 1D case, as first pointed out by \citet{Joyce2011,Benhaiem2013}. This is equivalent to assuming that relaxed objects become of constant size in the coordinate $r'= a^{1/3}x$ instead of the physical coordinate $r= ax$, as normally considered in the three-dimensional case \citep{Davis1977,Peebles:1980}, leading finally to $k\,P(k)\sim \mbox{const}.$ at large $k$ in our set-up with the initial power spectrum given by Eq.~(\ref{eq:initial_power_spec}). At low $z$, pure Zel'dovich predictions and post-collapse PT strongly underestimate $P(k)$, which is also expected in 3D. As found earlier by \cite{Taruya_Colombi2017}, results are considerably improved when employing adaptive smoothing, even in the stable clustering regime. In all the cases, as expected, post-collapse PT behaves slightly better than Zel'dovich approximation.

To further illustrate how various approximations perform compared to the $N$-body result, a portion of phase-space is shown at the same three redshifts on the bottom panels of Fig.~\ref{pk-fig} for one of the realizations. Since the Zel'dovich approximation does not account for back-reactions due to gravitational dynamics in the multi-stream regime, the Zel'dovich phase-space density gets increasingly stretched with time, resulting in less power at large $k$. A similar behaviour is seen for post-collapse PT, despite small scale corrections of the motion in the three-streams regime. The adaptive smoothing procedure drastically improves the visual agreement between theoretical modelling and the $N$-body simulation, at the cost of discontinuities of the phase-space sheet. As we mentioned in Sec.~\ref{sec:PCPT}, these discontinuities appear manifest at the transitions between single- and multi-stream flows or between two multi-stream flows corresponding to different halo mass scales. Nevertheless, the spatial extent of these regions is small enough, and thus the discontinuities are expected to have an impact on power spectrum mainly at small scales lying outside the plotted range. Yet, this impact is non-trivial in the sense that it represents, in practice, a significant source of noise on the response function, as we shall see below. Despite these discontinuities, the improvement brought by the adaptive smoothing algorithm is unquestionable and particularly striking for post-collapse PT, which provides a good description of the size of the haloes, summarizing their internal structure in a very rough yet reasonable way.


\subsection{Response function in one dimension}
\label{sec:respev}

\begin{figure*}
\centering
\includegraphics[width=12.7cm]{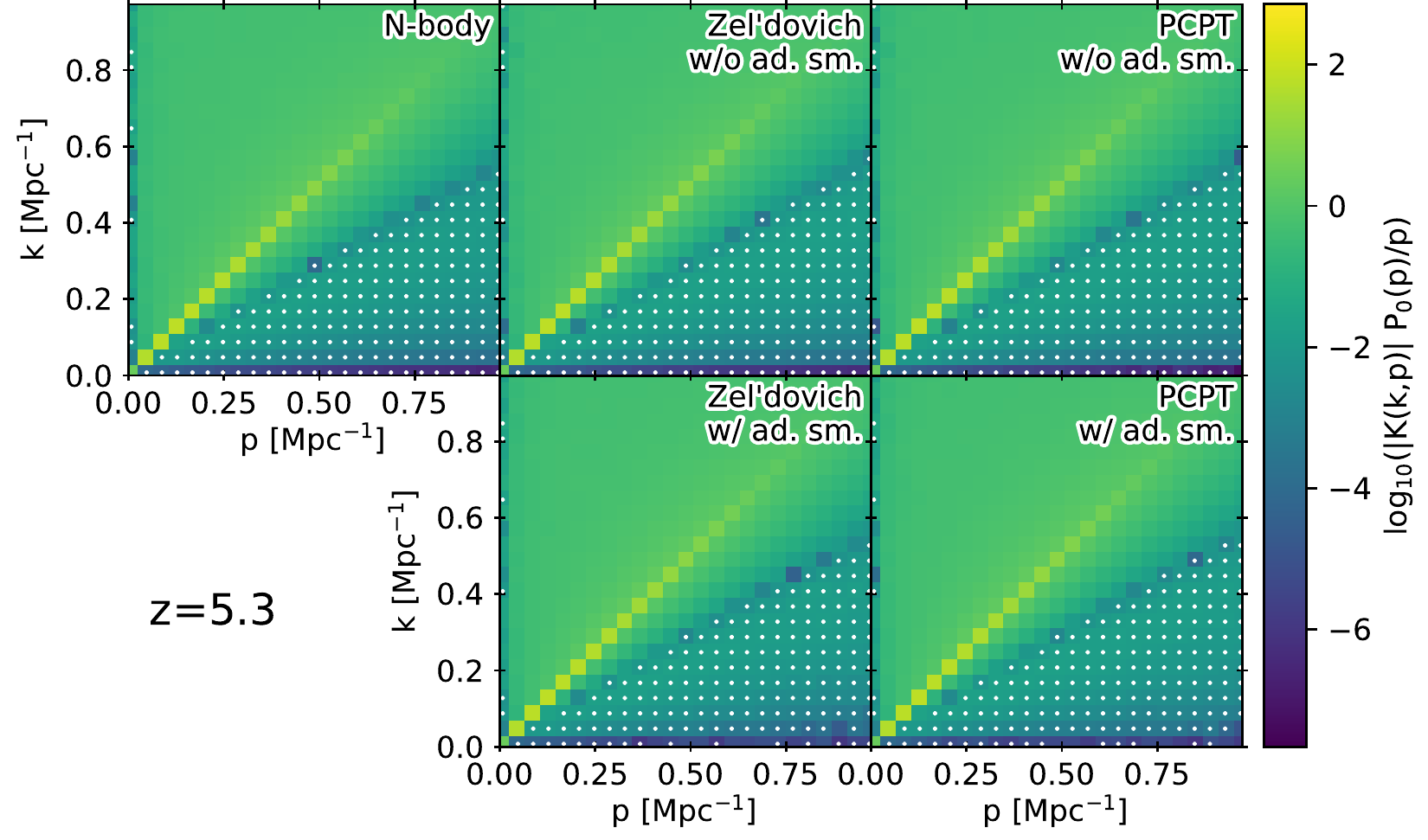}
\includegraphics[width=12.7cm]{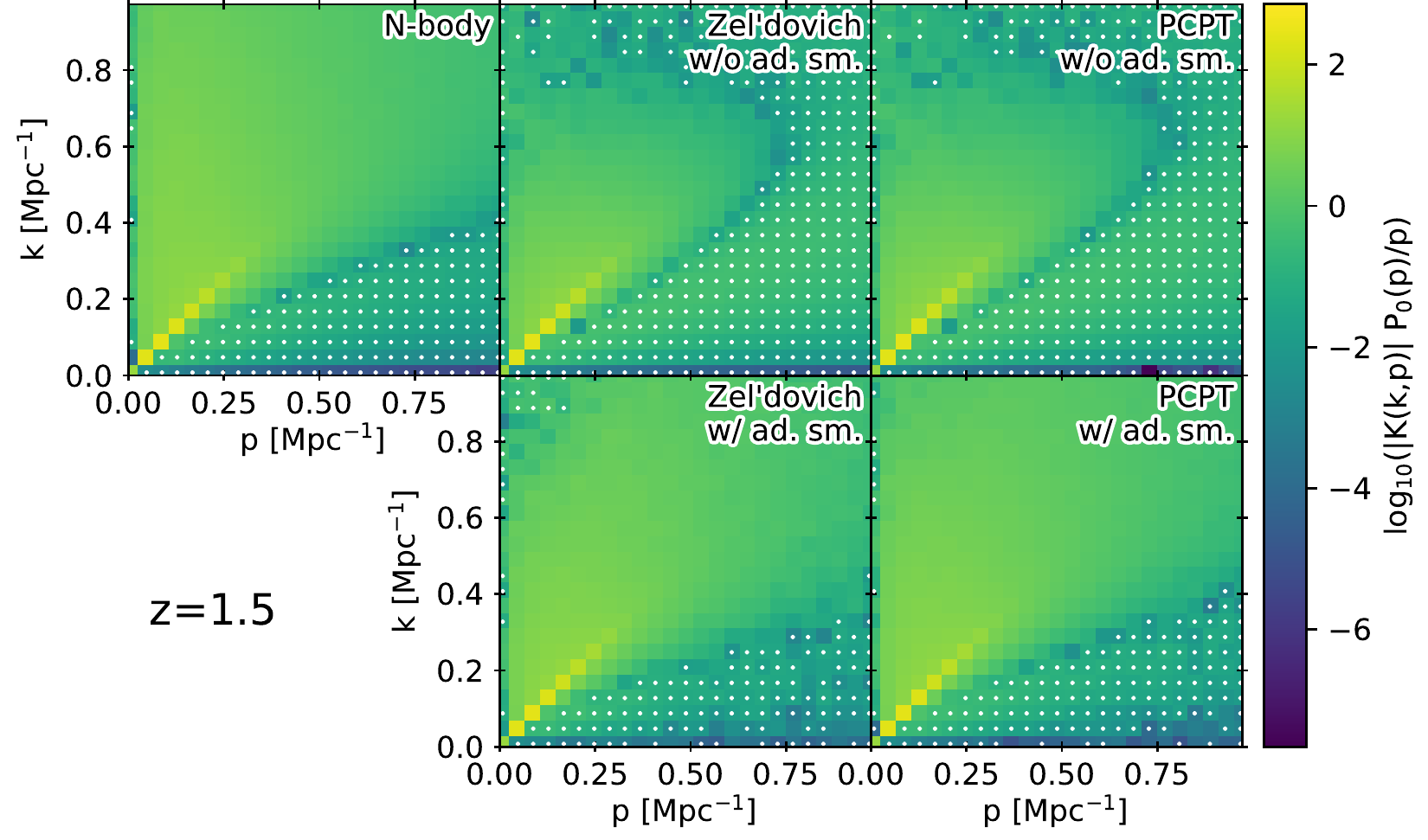}
\includegraphics[width=12.7cm]{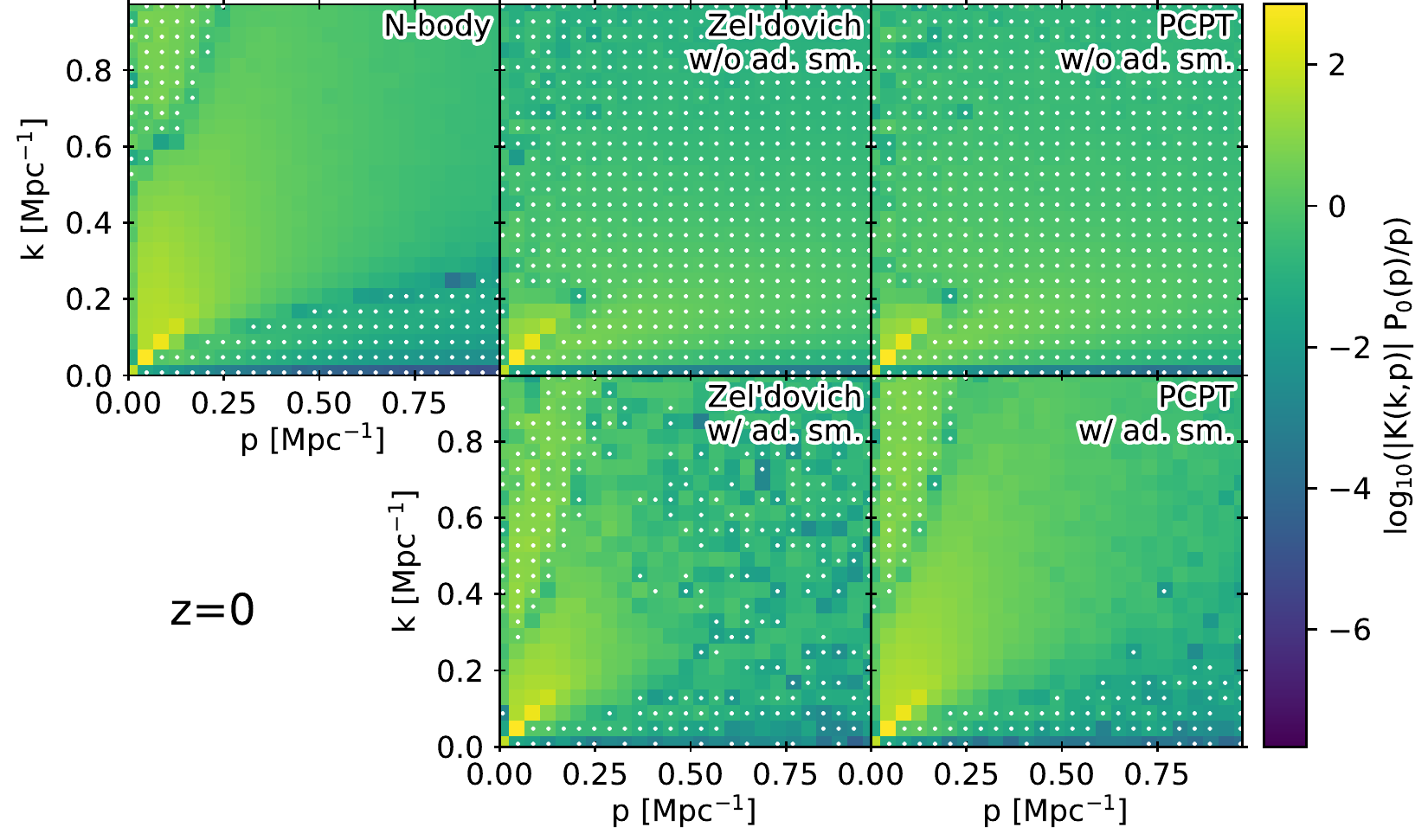}
\caption{Absolute value of the response function (colour scale) multiplied by $P_0(p)/p$ at three redshifts, using a logarithmic colour table for the representation. On each panel, white dots indicate negative values of the response function. Abbreviations used on each panel are explained in caption of Fig.~\ref{pk-fig}.}
\label{2D-fig}
\end{figure*}

Response functions for $N$-body, Zel'dovich and post-collapse PT solutions, with or without adaptive smoothing, are represented for all $k$'s on Fig.~\ref{2D-fig} at the three redshifts we consider. In each panel, the $25 \, \times \, 25$ values corresponding to the binning in $p$ and $k$ are sampling the $[0,1]$~Mpc$^{-1} \, \times \, [0,1]$~Mpc$^{-1}$ plane. The plotted results are the absolute values of the response function multiplied by $P_0(p)/p$, i.e., $|K(k,p)| P_0(p)/p$.

Consistently with the top left panel of Fig.~\ref{pk-fig}, response functions all look very similar at $z=5.3$. They exhibit a peak at $k$=$p$, surrounded by tails. Note that the response function takes positive values at $k>p$, while it becomes negative at $k<p$ (lower-right triangle region in each panel). Although the actual tails away from the peak are highly suppressed, indicating little mode coupling, the resultant behaviours are rather contrasted with the Eulerian linear theory prediction, $K(k,p;z) \propto \delta_{\rm D}(k-p)$, where $\delta_{\rm D}$ is the Dirac delta function. On the other hand, at $z=1.5$, significant differences between various solutions appear, especially for wavenumbers $\gtrsim 0.7$~Mpc$^{-1}$, although the overall trends in $N$-body simulations look similar to those at $z=5.3$. Zel'dovich and post-collapse PT solutions without adaptive smoothing exhibit similar deviations from the $N$-body results, with negative values in a larger area of the lower-right triangle region. That is, these two solutions predict a significant amount of coupling between small- and large-scale Fourier modes, compared to the $N$-body results. This is likely due to a bad description of the multi-streaming regime, as previously observed. Indeed, adaptive smoothing improves the results, especially for post-collapse PT. At $z=0$, the situation becomes even worse, and pure Zel'dovich dynamics and post-collapse PT perform very poorly. By contrast, the improvement brought by adaptive smoothing is drastic. One however notices that results obtained with adaptive smoothing are noisy (especially for Zel'dovich dynamics), which is likely due to the discontinuities of the phase-space sheet introduced by the implemented procedure, as previously discussed, in Secs.~\ref{sec:PCPT} and \ref{sec:powev} (bottom panels of Fig.~\ref{pk-fig}).

\begin{figure*}
\centering
\resizebox{\hsize}{!}{
\begin{tabular}{c}
\includegraphics{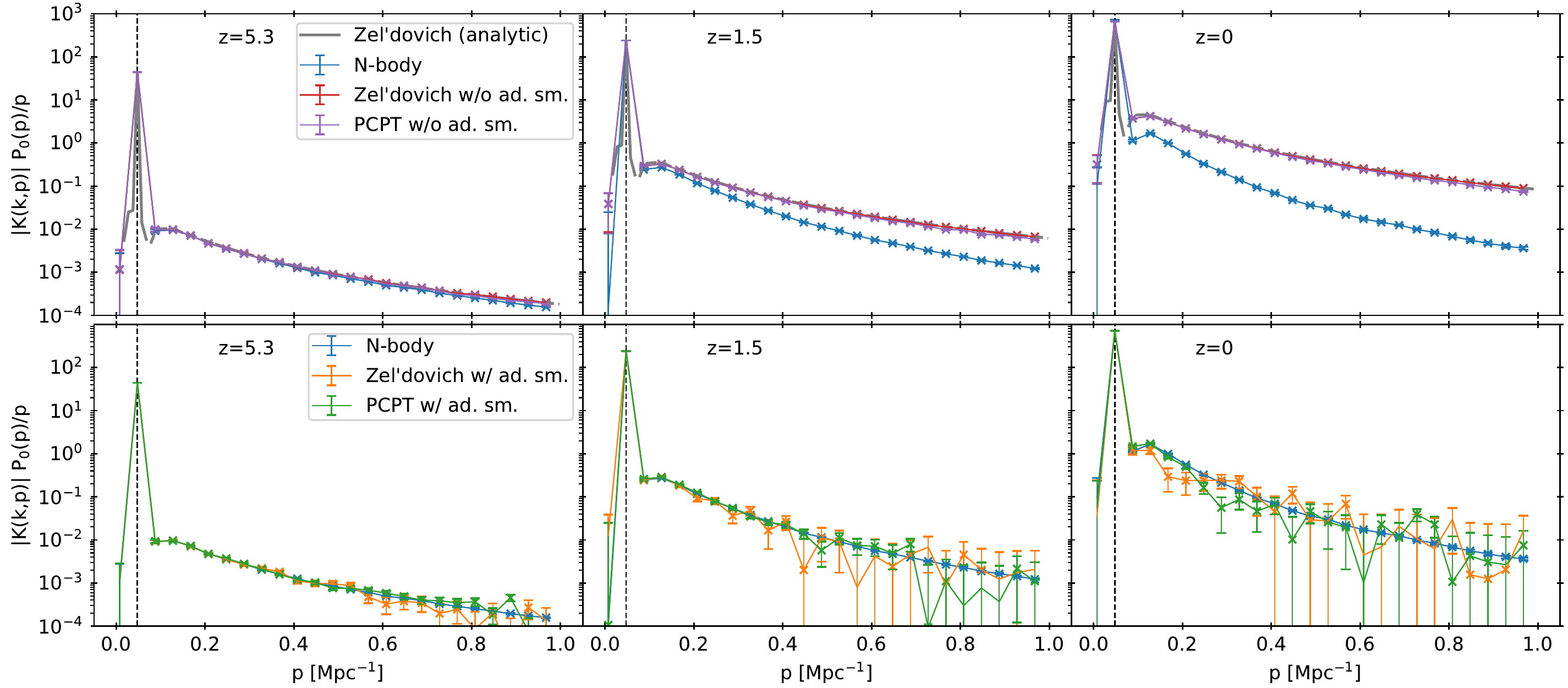} \\
\includegraphics{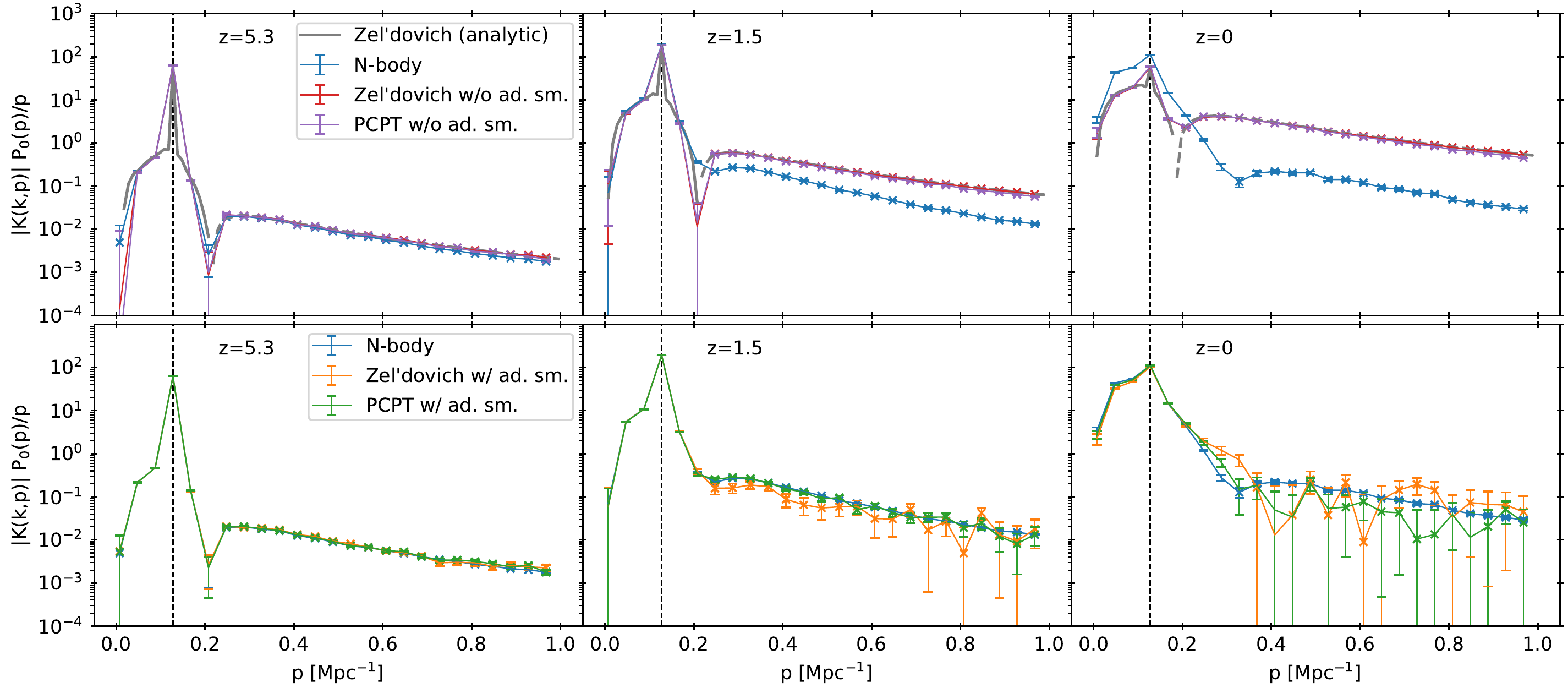} \\
\includegraphics{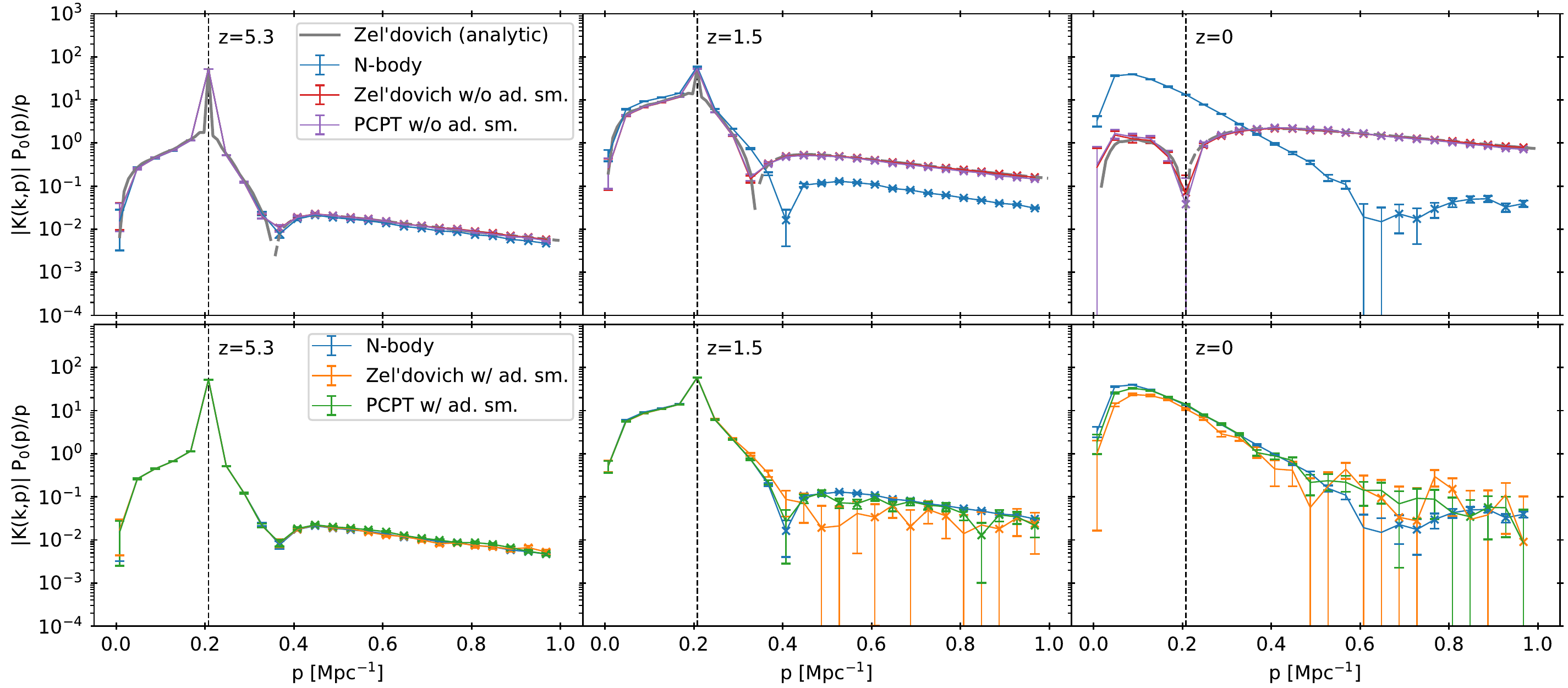} \\
\end{tabular}}
\caption{Absolute value of $K(k,p;z)P_0(p)/p$ as a function of $p$ for different values of $k$ (indicated by a dashed vertical line) at the three redshifts we consider. From top to bottom, $k=0.0475, 0.1275, 0.2075$~Mpc$^{-1}$. Small crosses indicate negative values and error bars represent the statistical error given by $\sigma/\sqrt{N_{\rm sim}-1}$ with $\sigma$ the standard deviation obtained from the estimator (\ref{eq:response_estimator}) for the number $N_{\rm sim}$ of pairs of simulations we used. Abbreviations used on each panel are explained in caption of Fig.~\ref{pk-fig}.}
\label{1D-fig}
\end{figure*}

\begin{figure*}
\centering
\resizebox{\hsize}{!}{
\begin{tabular}{c}
\includegraphics{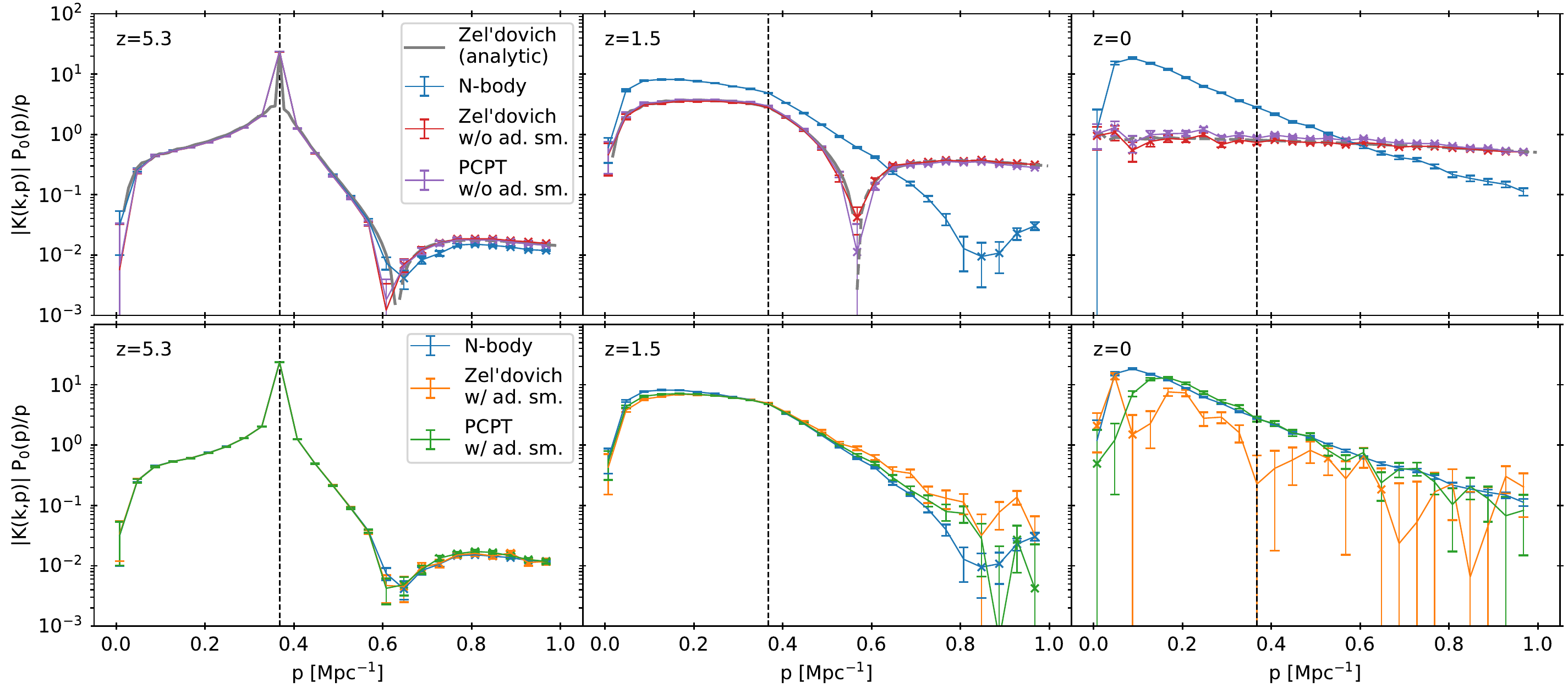} \\
\includegraphics{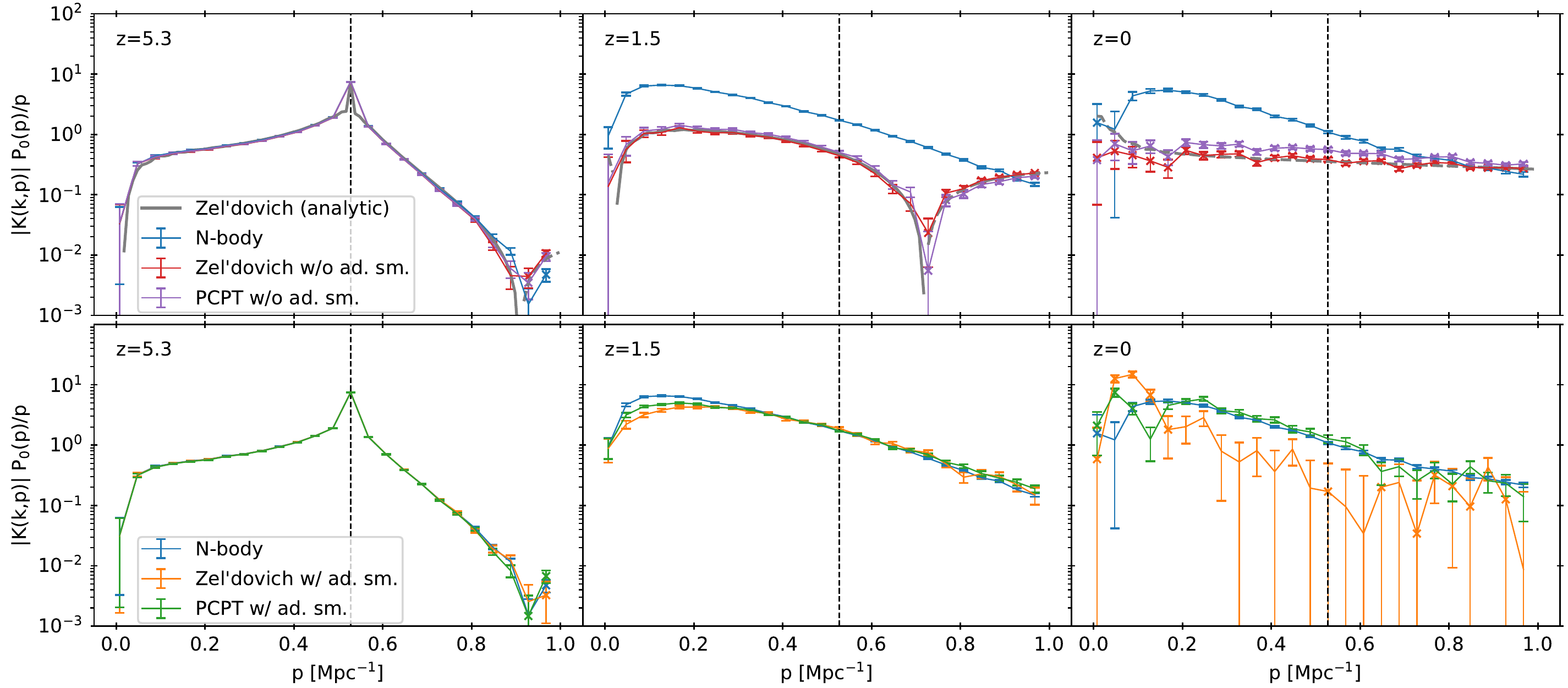} \\
\includegraphics{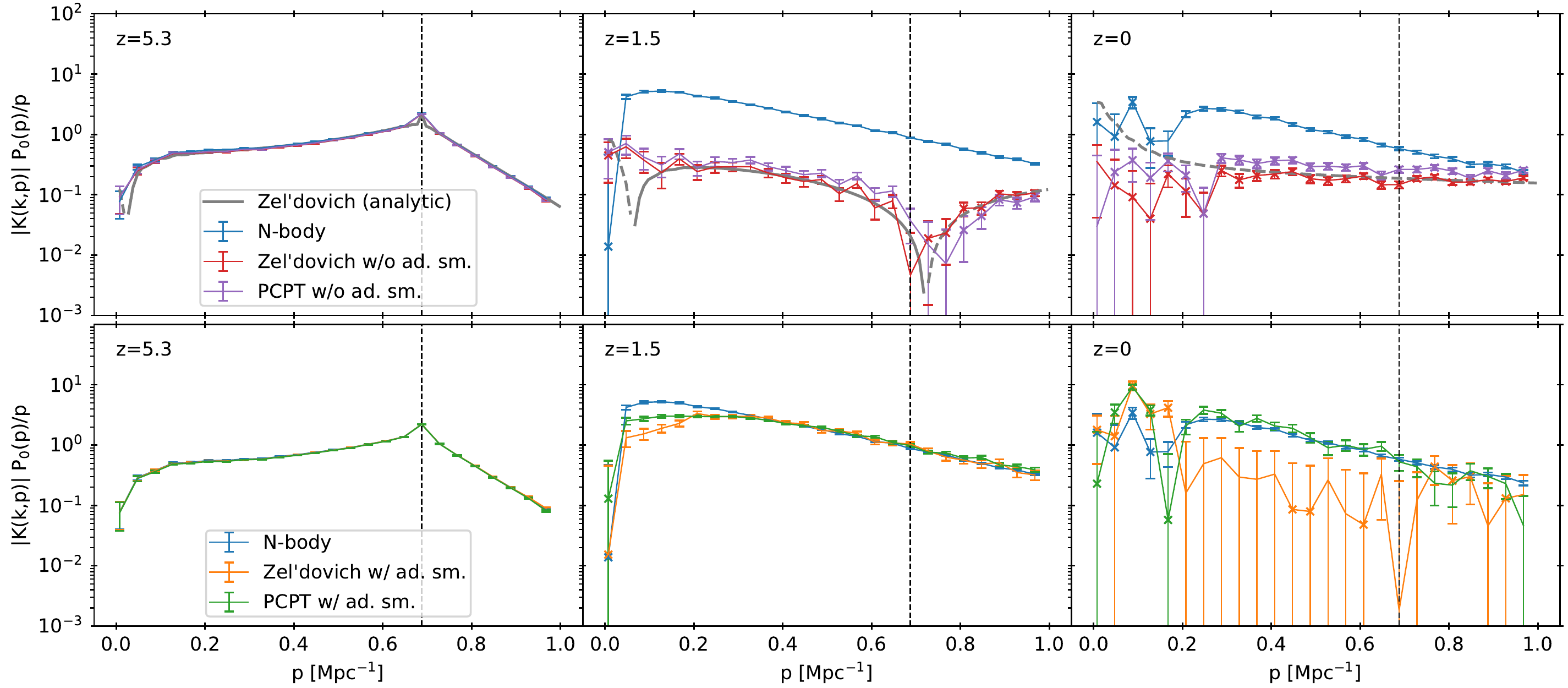}
\end{tabular}}
\caption{Same as Fig.~\ref{1D-fig} for larger modes $k=0.3675, 0.5275, 0.6875$~Mpc$^{-1}$.}
\label{1Dbis-fig}
\end{figure*}

To discuss various regimes in more detail, we now examine Figs.~\ref{1D-fig} and \ref{1Dbis-fig}, which show the response functions $K(k,p;z)$ for a few fixed values of $k$. Despite the fact that the 1D power-spectrum is damped instead of enhanced in the non-linear regime, the visual comparison of these figures to Fig.~2 of \citet{Nishimichi_etal2017} shows that $N$-body measurements of the response function are similar in many respects to what was obtained in 3D \citep{Nishimichi:2014rra,Nishimichi_etal2017}. There is always a decrease in the response function when approaching $p=0$, consistently with the cancellation of the infrared contributions expected from Galilean invariance \citep{peloso2013}, as well as a change of sign of $K(k,p)$ at large $p$. As in 3D, the peak around $k=p$ can be suppressed along with the appearance of a local maximum at $p < k$ (see for example the lower panels of Fig.~\ref{1D-fig} which treat the $k=0.2075$~Mpc case). Detailed inspection of this phenomenon suggests however that the peak structure of the response function is suppressed faster in 1D than in 3D.

We now discuss Zel'dovich and PT predictions without adaptive smoothing, by examining, on Figs.~\ref{1D-fig} and \ref{1Dbis-fig}, the first rows in each group of 6 panels. The grey curves (almost indistinguishable from the red curves) correspond to the analytic Zel'dovich response function obtained from Eq.~(\ref{eq:kernel_ZA_1Dmain}) for $k$ and $p$ bins centred at $ \pi/L + (0.5+n) \Delta p $, $n={0,\ldots, 99}$, with the same values of $L=1260$~Mpc and $\Delta p=0.01$~Mpc$^{-1}$ as for the other measurements. Positive and negative values of the analytic results are shown as solid and dashed lines, respectively. The analytic prediction follows very closely the red curves which correspond to the statistical averaging over the Zel'dovich solutions obtained directly from the $N$-body initial conditions (this match is not expected to be perfect because there is no cut-off introduced above $k_{\rm c}$ in the semi-analytical computation). When $z \le 1.5$, the Zel'dovich and post-collapse PT solutions strikingly differ from the $N$-body solution. For $k=0.0475$~Mpc$^{-1}$ or $k=0.1275$~Mpc$^{-1}$ (first and third rows of Fig.~\ref{1D-fig}), the coupling of small scales with these large scales is significantly lower in amplitude in the $N$-body case. For $k=0.1275$~Mpc$^{-1}$ the peak around $k=p$ at $z=0$ is higher and wider for the $N$-body solution, indicating a less efficient coupling between modes in this regime for the Zel'dovich or the post-collapse PT approximations. At larger $k$ and intermediate to low redshifts, we showed in the previous section that the power-spectrum is damped at small scales by Zel'dovich and post-collapse PT dynamics, due to the artificial stretching of multi-flows regions. As illustrated by the bottom panels of Fig.~\ref{1D-fig} and by Fig.~\ref{1Dbis-fig}, this has the effects of flipping the sign and flattening the shape of corresponding response functions, which become totally inconsistent with the $N$-body result.

Examining now the second rows in each group of six panels of Figs.~\ref{1D-fig} and \ref{1Dbis-fig}, we confirm again that adaptive smoothing remedies the flaws of Zel'dovich and post-collapse PT. It provides response functions much closer to the $N$-body results, especially when post-collapse PT is used. As already discussed above, measurements with adaptive smoothing are however subject to significant noise, particularly Zel'dovich approximation. Other discrepancies are noticeable, e.g. for post-collapse PT at low $p$'s in the bottom right panel of the first group of six panels Fig.~\ref{1Dbis-fig}, that are not easily explainable by fluctuations of the noise or obvious defects of the adaptive smoothing procedure, but that are consistent with the fact that the power-spectrum given by post-collapse PT slightly underestimates that of the simulations at large $k$ and low redshifts (see Fig.\ref{pk-fig}).

\section{Three-dimensional cosmology}
\label{sec:implications}
The results of the last section show that adaptive smoothing used with post-collapse PT helps a lot reproducing $N$-body measurements, indicating the importance of (clever) coarse-graining in order to account for the multi-stream regime. While post-collapse PT dynamics is not straightforward to resolve in 3D, there still exist ways to incorporate some regularization of particle trajectories after shell crossing. PINpointing Orbit Crossing Collapsed HIerarchical Objects ({\tt PINNOCHIO};\footnote{\tt http://adlibitum.oats.inaf.it/monaco/pinocchio.html} \citealt[][]{2002MNRAS.331..587M}) is one of such techniques, based on Lagrangian perturbation theory (LPT) and a modified version of the peak-patch theory \citep{1996ApJS..103....1B}.

\subsection{The PINOCCHIO algorithm}
\label{subsec:pino}

\texttt{PINOCCHIO} was originally developed to simulate merger trees of dark matter haloes and was often used as a quick algorithm to produce mock galaxies. In its first step, it adopts an ellipsoidal collapse model based on Lagrangian perturbation theory and computes the time when the first axis collapses at various smoothing scales. Then, the earliest collapsing time over different smoothing scales is recorded as the collapse time for each mass element. Next, it scans over time and progressively connects nearby collapsed particles by a friend-of-friend like algorithm to form a filamentary network and a merger tree in a manner resembling the hierarchical clustering of structures. This way, the fate of all the collapsed particles is determined; particles are either marked as belonging to a given halo or marked as ``filament'' particles. In its latest version, which we use in this work (V4; \citealt[][]{Munari17}), \texttt{PINOCCHIO} adopts second-order LPT (2LPT) to construct the merger tree and third-order LPT (3LPT) to displace the haloes centres once they have been identified.

While the main usage of the code is to simulate haloes, it also computes the distribution of particles. These latter are displaced from an initial regular lattice using LPT, except those classified to constitute a halo, which the code relocates to form a sphere with the Navarro-Frenk-White (NFW; \citealt[][]{NFW}) radial density profile around the halo centre. Thus, it can be viewed in our context as an LPT implementation with regularized dynamics after forming a halo. Indeed, the application of multi-scale filters closely resembles our adaptive smoothing technique as both methods rely on the calculation of dynamical times to decide on the local smoothing scale. The main differences lie in the actual calculation of the dynamical time and the way multi-streams are treated. In our 1D procedure based on post-collapse PT, the dynamical time relates to next crossing instead of collapse. In addition to relying on 2LPT displacements combined with a friend-of-friend algorithm to account for mergers, \texttt{PINNOCHIO} uses NFW profiles to describe the multi-stream regime inside haloes, while, in 1D, we represent haloes at the coarse level with ``S'' shapes in phase-space and use the Lagrangian size given to these coarse haloes by post-collapse PT to account for mergers in Lagrangian space.

\subsection{Set up of numerical experiments}
\label{subsec:3dsetup}
We now describe the numerical experiments we performed to study the response function in three dimensions. In order to cover the dynamic range of interest, we consider wavenumbers spanning the interval $[0.003,1.09]\,h\,\mathrm{Mpc}^{-1}$ (with $h=H_0/100$), using logarithmic bins corresponding to a factor $\sqrt{2}$ between two successive values of $k$. We employ this binning scheme for both $k$ and $p$ to form a $17\times17$ matrix to sample the function $K(k,p)$. Note that, in the subsequent analyses, the first two bins corresponding to the two largest scales will be covered only by the $N$-body simulations and not by the theoretical models. This is related to the fact that a smaller cosmological volume was used to study LPT and {\tt PINNOCHIO}, as detailed below.

We run \texttt{PINOCCHIO} in a flat $\Lambda$CDM cosmology with parameters based on the five-year observations of the WMAP satellite \citep{Komatsu:2008hk}.
We consider a periodic cube of size $L=1024\,h^{-1}\mathrm{Mpc}$ with $N_{\rm p}=512^3$ mass elements. For convergence study, we also consider $N_{\rm p}=384^3$, $640^3$, $768^3$, $896^3$ and $1024^3$ mass elements with the same box size. We run $100$ random realizations to study the matter power spectrum for each of the $6$ different resolutions. We also create in total of $3\,000$ realizations ($=100$ pairs of simulations with different random seeds for each bin) with perturbed linear power spectra for the default setting of $512^3$ mass elements to study the response function. Following \citet{Nishimichi:2014rra}, we consider a $\pm 1\%$ modulation of the linear power spectrum in the interval $[p_{\rm min},p_{\rm max}[=[p_i,p_{\rm i+1}[$ for one of the 15 $p$ bins, where $i$ is the bin number. In addition, $800$ supplementary simulations ($=100$ seeds but only for the $4$ largest $p$ bins) are done for each of the $5$ different settings for the mass resolution. We implement this in \texttt{PINOCCHIO} by modifying the source code to accept the location of the $p$ bin specified by $(p_\mathrm{min},p_\mathrm{max})$ and the size of the perturbation in the linear power spectrum as the inputs in the parameter file.

For comparison, we generate LPT realizations at different orders [first = Zel'dovich approximation (ZA), second = 2LPT and third = 3LPT], with the same setting as the default calculation of \texttt{PINOCCHIO}. For consistency, we use the functions implemented in the \texttt{PINOCCHIO} package for this exercise. This time, we only consider the default setting of $512^3$ mass elements. Again, we run $100$ realizations per LPT order to calculate the matter power spectrum, and $300$ perturbed simulations ($=10$ seeds for $15$ $p$ bins) for the response function.

Finally, to assess the predictions of the models described above, we perform $N$-body simulations with $N=1024^3$ particles in a comoving periodic cube of size $L=2048\,h^{-1}\mathrm{Mpc}$ with the public Tree-PM code \texttt{Gadget2}~\citep{Springel:2005mi}, starting from 2LPT initial conditions generated with a code originally developed in \citet{Nishimichi:2008ry} and parallelized in \citet{Valageas:2010yw}. We generate $10$ random realizations to study the matter power spectrum and $68$ simulations from perturbed linear power spectra ($2$ seeds for $17$ $p$ bins) to study the response function.

We consider snapshots at $z=3, 2, 1$ and $0.5$ for all the numerical experiments described above. Power spectra are measured using Fast Fourier Transform on $1024^3$ grid points after mass assignment with the Cloud-in-Cell interpolation algorithm \citep{1988csup.book.....H}.
\subsection{Matter power spectrum in three dimensions}
\label{subsec:3dpk}

\begin{figure}
\includegraphics[width=8cm]{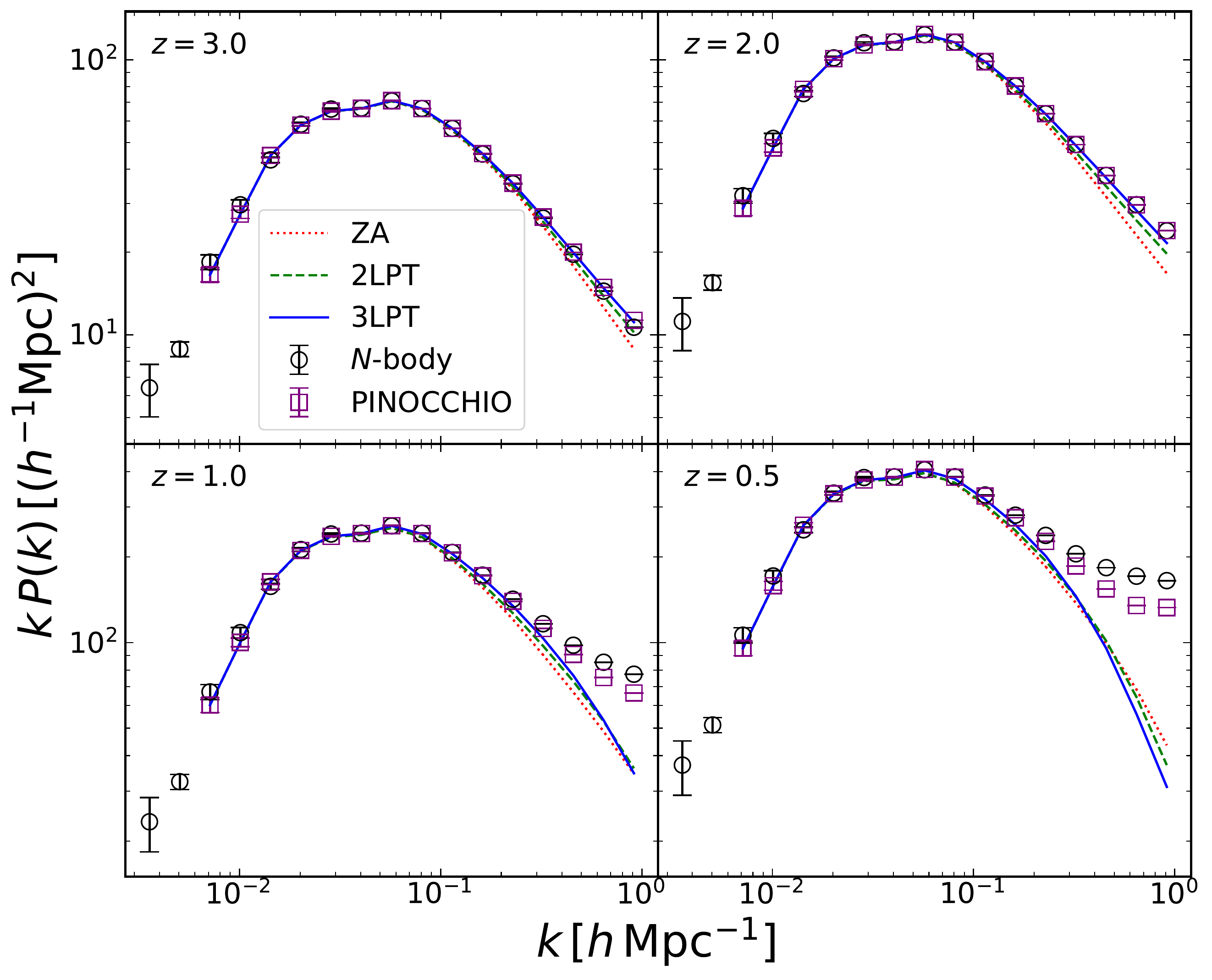}
\caption{Matter power spectrum from LPT at different order and \texttt{PINOCCHIO} compared with $N$-body simulations. The plotted results are the power spectra multiplied by $k$, i.e., $k\,P(k)$. }
\label{3Dpower}
\end{figure}

A marked difference between finite order LPT in 3D and Zel'dovich dynamics in 1D is that the former is only approximate while the latter is exact up to shell-crossing.
Before testing the response function, we thus first check the ability of LPT and \texttt{PINOCCHIO} to predict the matter power spectrum.

Figure~\ref{3Dpower} shows the power spectra for the four different epochs which we consider. For ease of comparison, the measured power spectra are multiplied by $k$. The results from the $N$-body simulations are represented by the circles with error bars. The corresponding predictions of LPT are depicted by lines of different types as indicated on the upper left panel, while the squares stand for \texttt{PINOCCHIO}. When increasing perturbative order, the agreement of LPT with $N$-body simulation slightly improves at high redshifts. However, this improvement remains small and takes place only in a finite range of wavenumbers. Indeed, going to higher order worsens the results at large $k$ for $z=0.5$. This suggests that no matter how high an order we may reach in the LPT calculation, the result may never converge to the $N$-body result on non-linear scales. In other words, even if we could obtain a full-order, supposedly exact, perturbative solution, it seems certain that we would not fully capture the non-linear dynamics of the cosmic fluid, because LPT is based on dynamical equations valid only in the single-stream regime.

Then, we can observe on Fig.~\ref{3Dpower} that the \texttt{PINOCCHIO} result remains close to the $N$-body simulations, even at low redshifts.
We have tested different numbers of particles for the \texttt{PINOCCHIO} realizations [$N=384^3$, $512^3$ (our default, shown on the figure), $640^3$, $768^3$, $896^3$ and $1024^3$ particles with the same box size], but the results look almost unchanged from visual inspection of this logarithmic plot. While the agreement with the $N$-body simulation is not perfect, the improvement brought by \texttt{PINOCCHIO} is remarkable, making it a good approximate model. Since the difference between \texttt{PINOCCHIO} and pure 3LPT lies only in the treatment of the ``halo particles'', i.e., the undesired particle trajectories in LPT after the formation of haloes are regularized to form an NFW sphere, these result already suggests that dynamics after shell-crossing plays an important role to get the power spectrum right.

\subsection{Response function in three dimensions}
\label{subsec:3dresponse}

\begin{figure*}
\includegraphics[width=8cm]{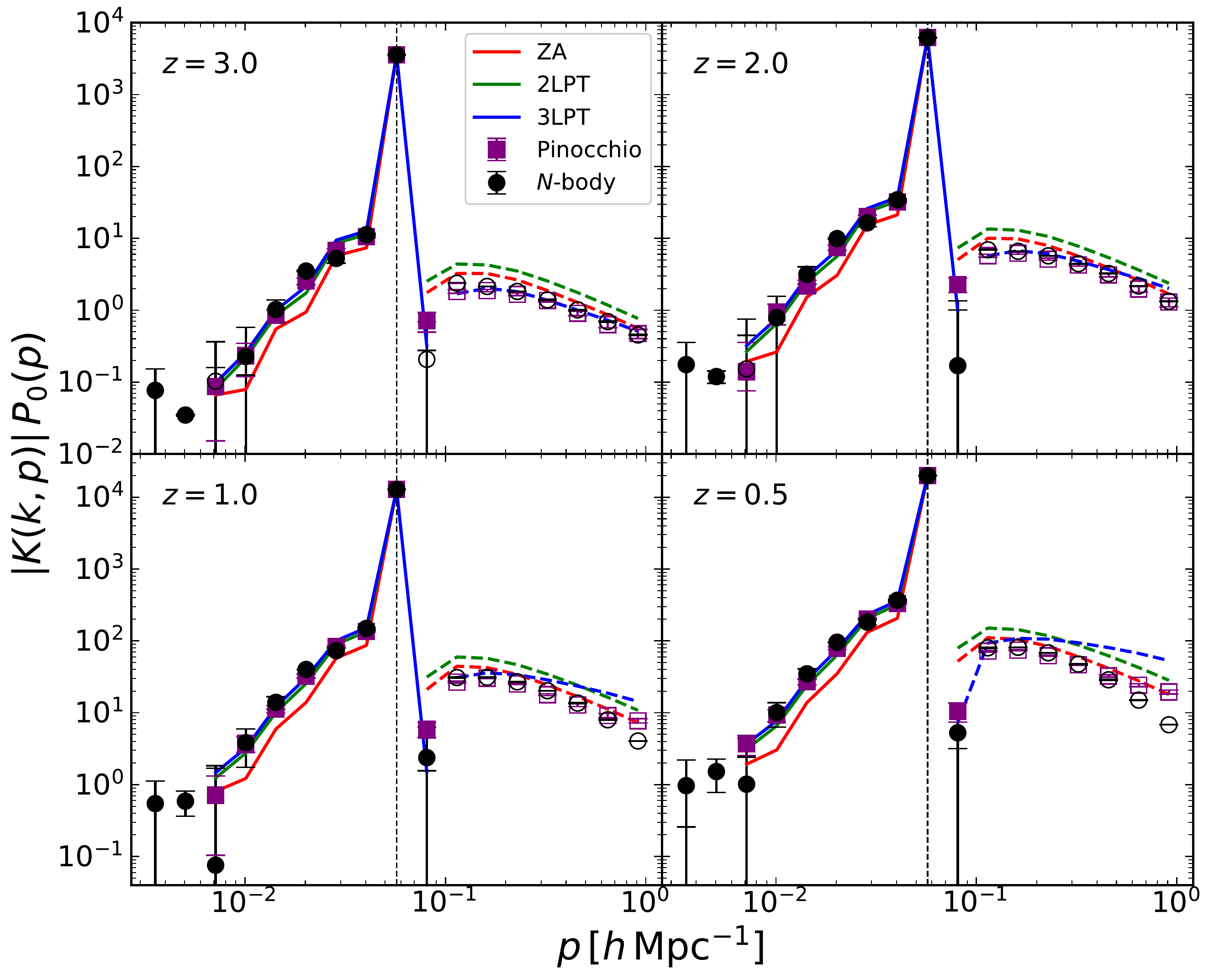}
\includegraphics[width=8cm]{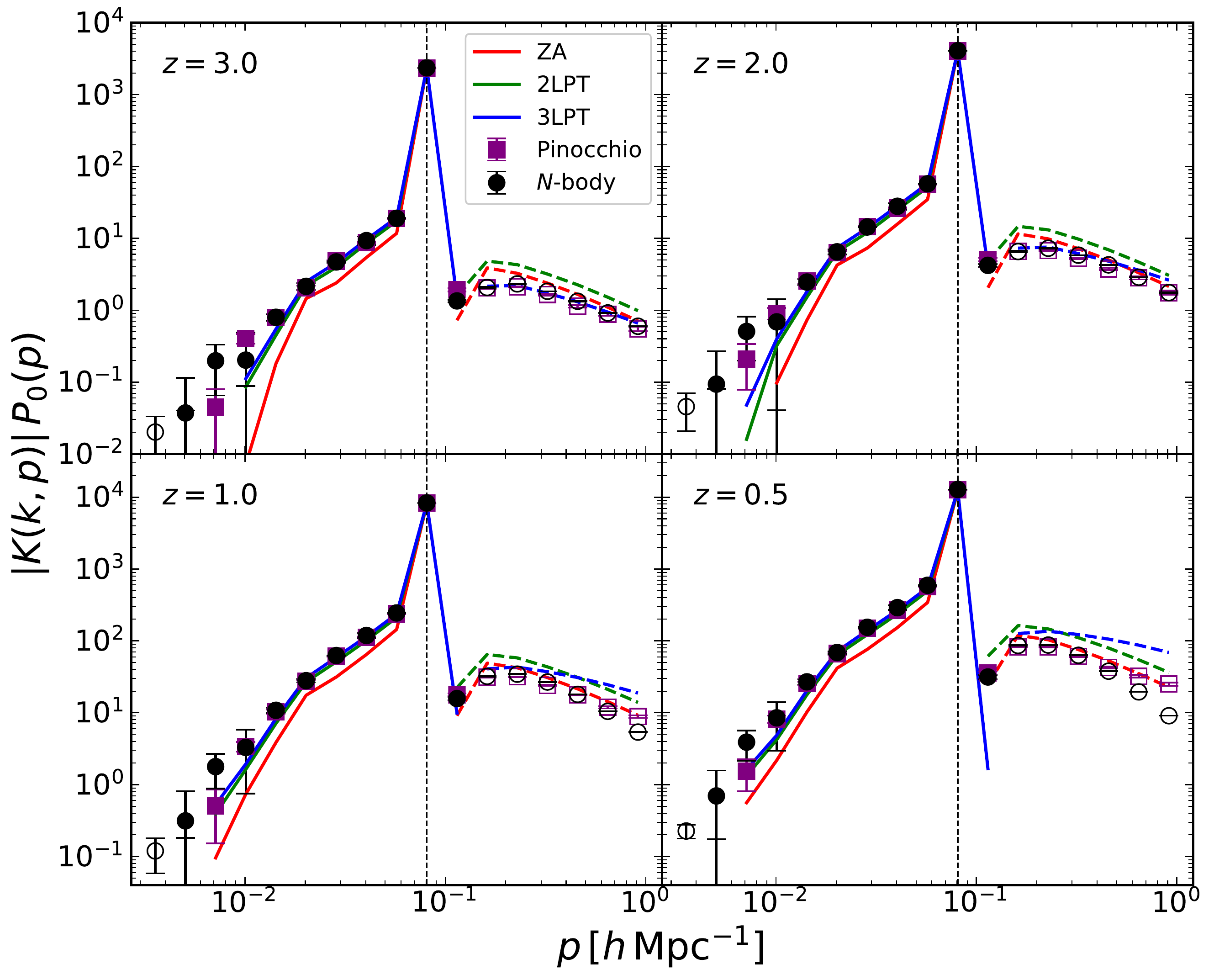}
\includegraphics[width=8cm]{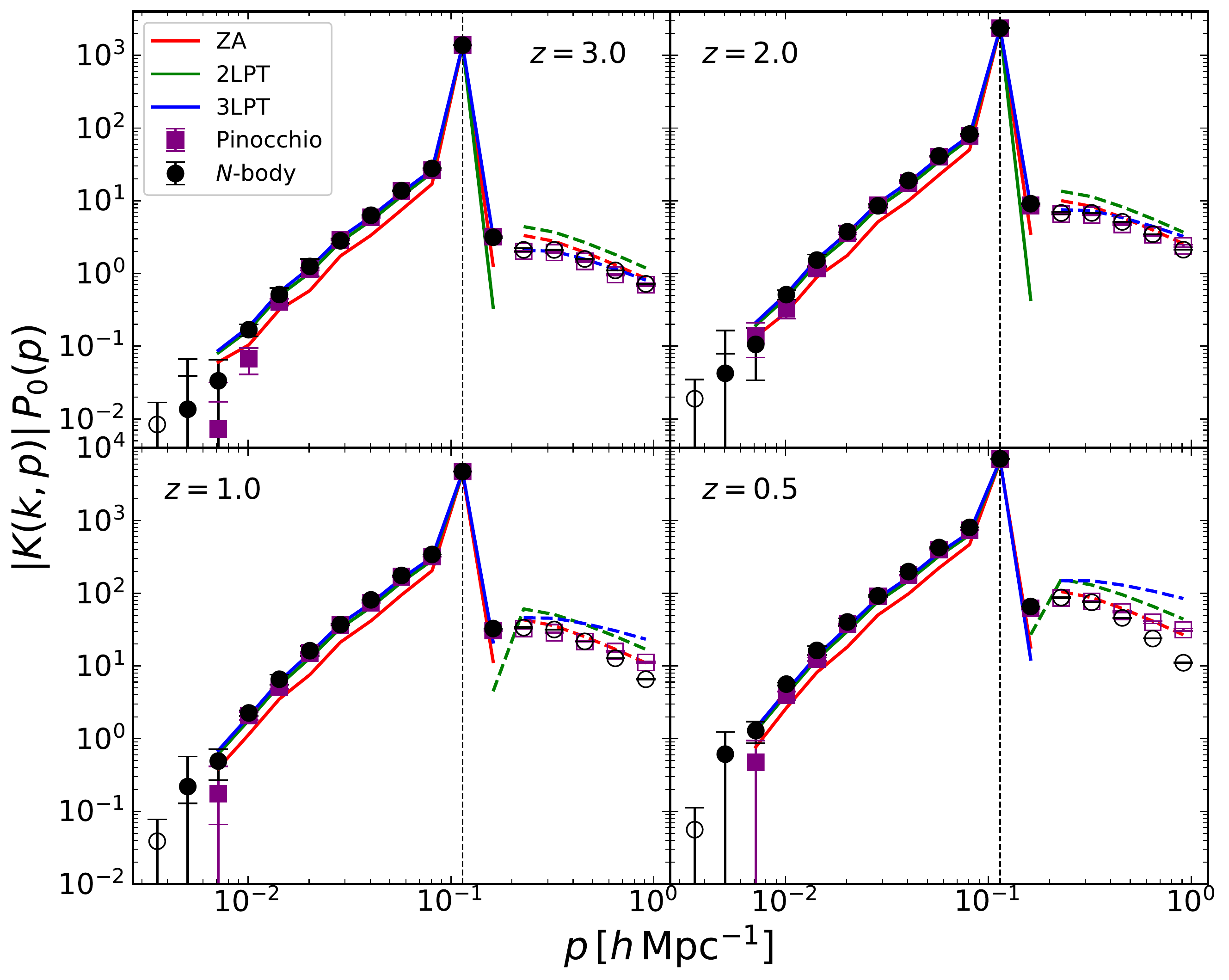}
\includegraphics[width=8cm]{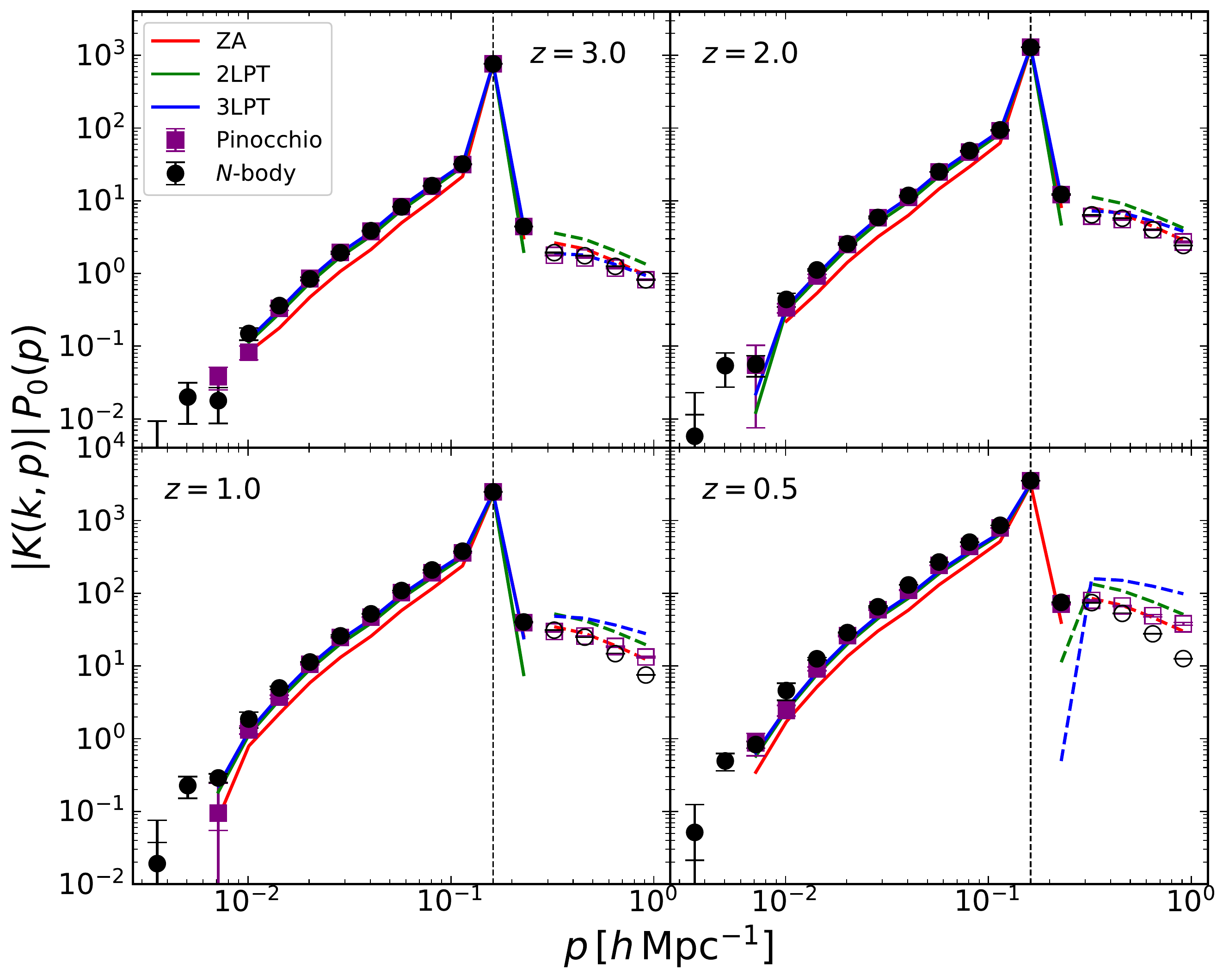}
\caption{Response function in three dimensions from LPT (lines) and \texttt{PINOCCHIO} (squares), compared with $N$-body simulations (circles). On each panel, we plot $|K(k,p)|\,P_\mathrm{lin}(p)$ as a function of $p$ for a fixed value of $k$ probing the weakly non-linear regime (from left to right and top to bottom, $k=0.057$, $0.081$, $0.114$ and $0.161 h$ Mpc${}^{-1}$). A positive (negative) value of $K(k,p)$ is depicted by a filled (open) symbol. Likewise, the lines are solid (dashed) when the response function is positive (negative). The vertical dashed lines are drawn at $p=k$.}
\label{3DResponse}
\end{figure*}

\begin{figure*}
\includegraphics[width=8cm]{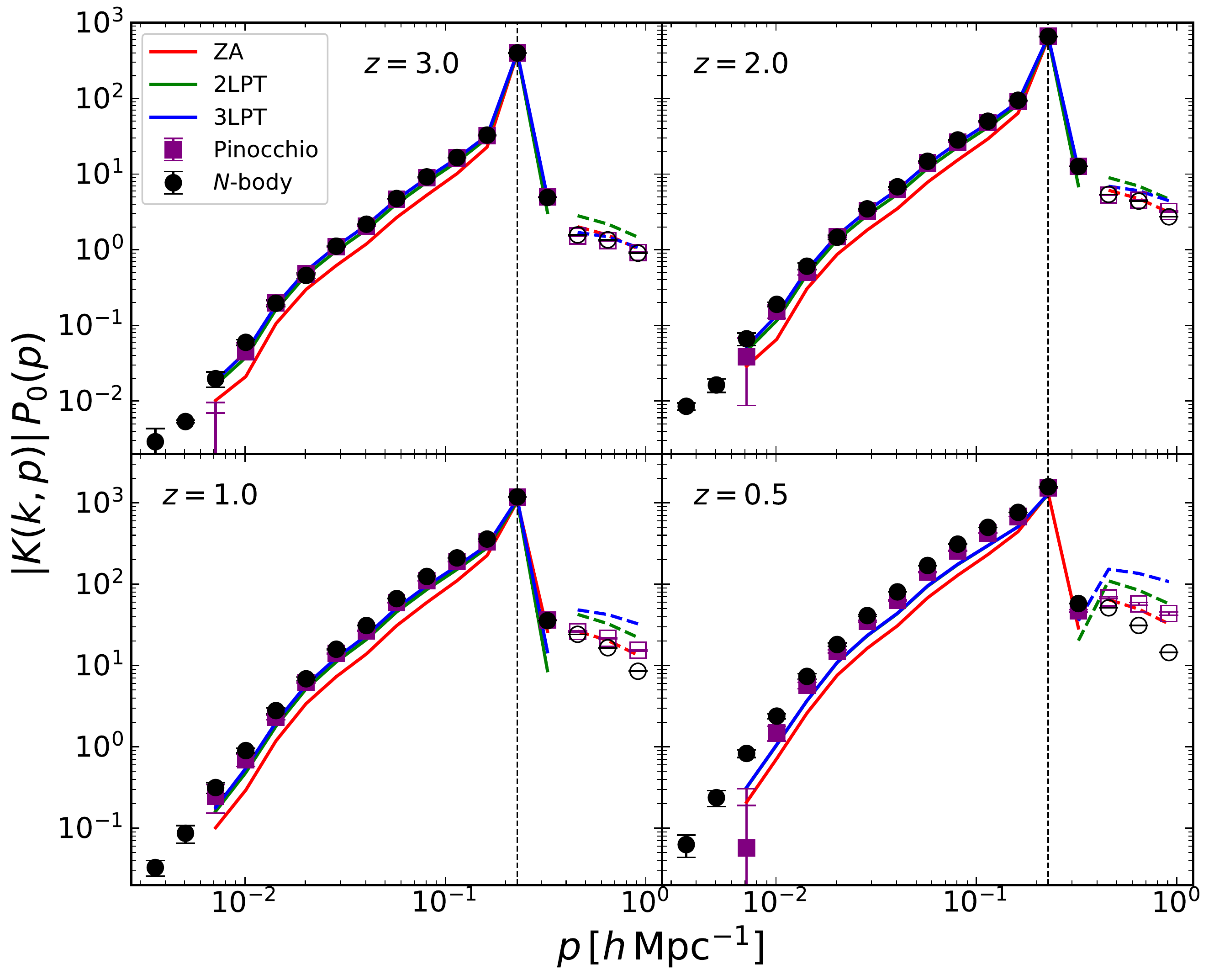}
\includegraphics[width=8cm]{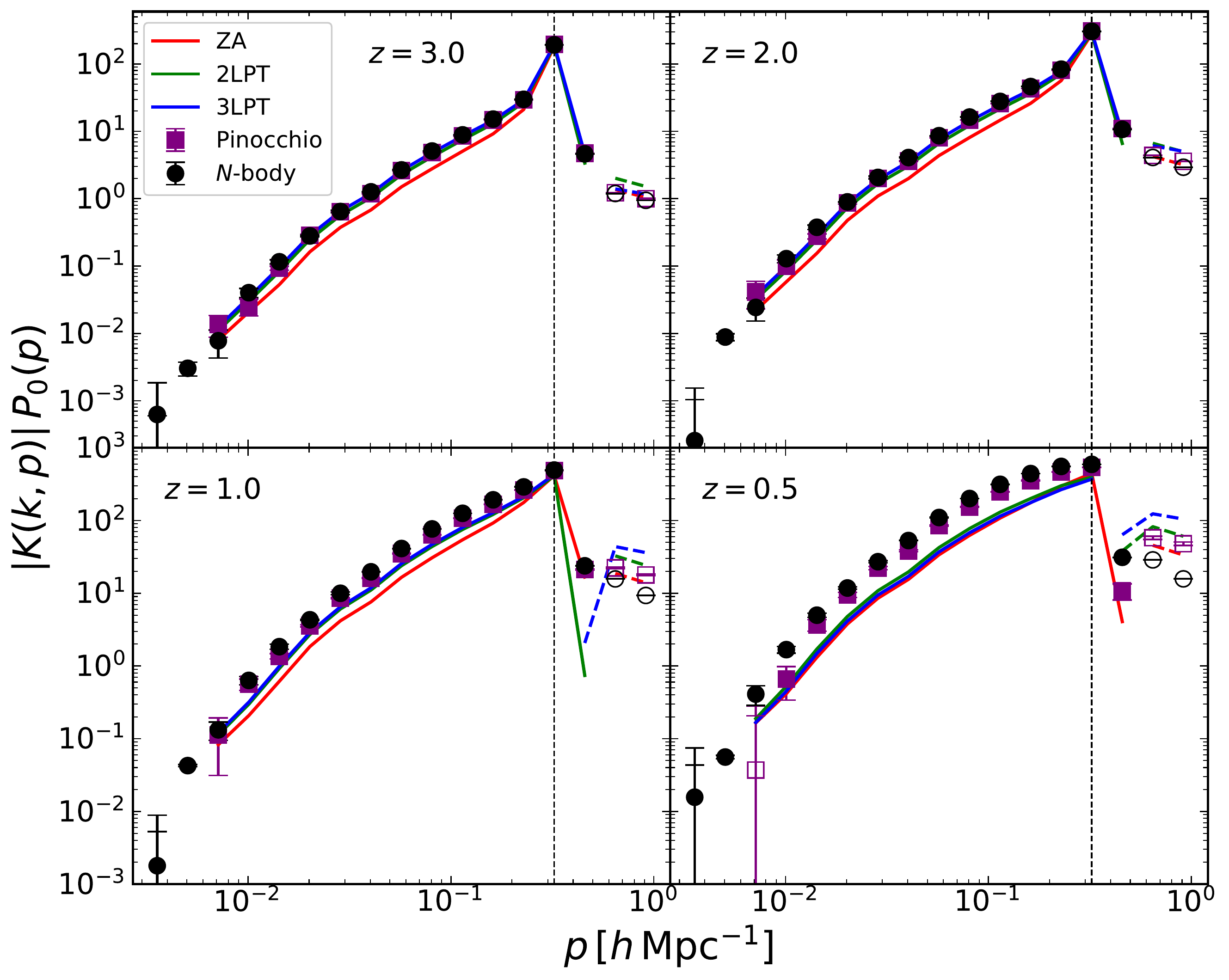}
\includegraphics[width=8cm]{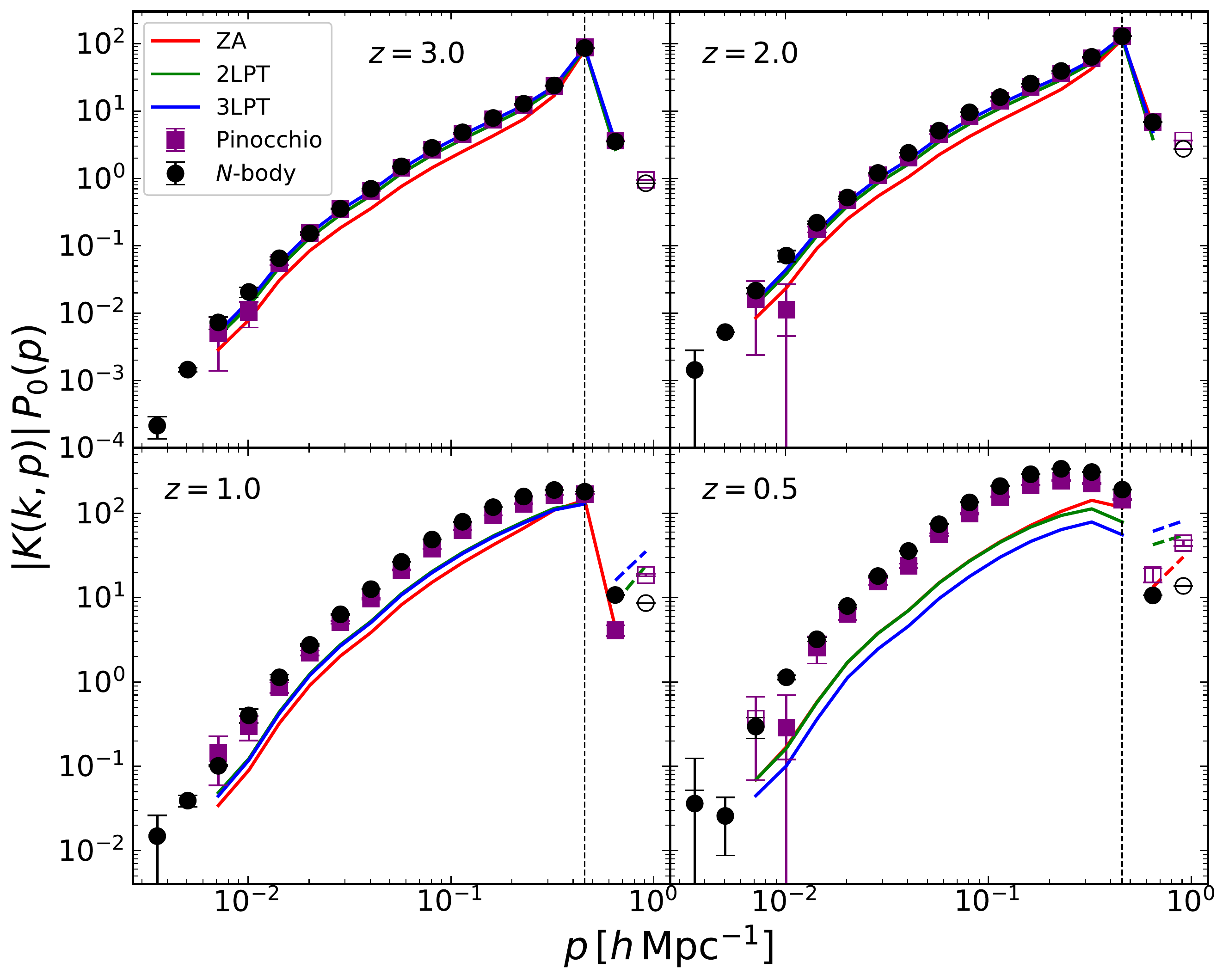}
\includegraphics[width=8cm]{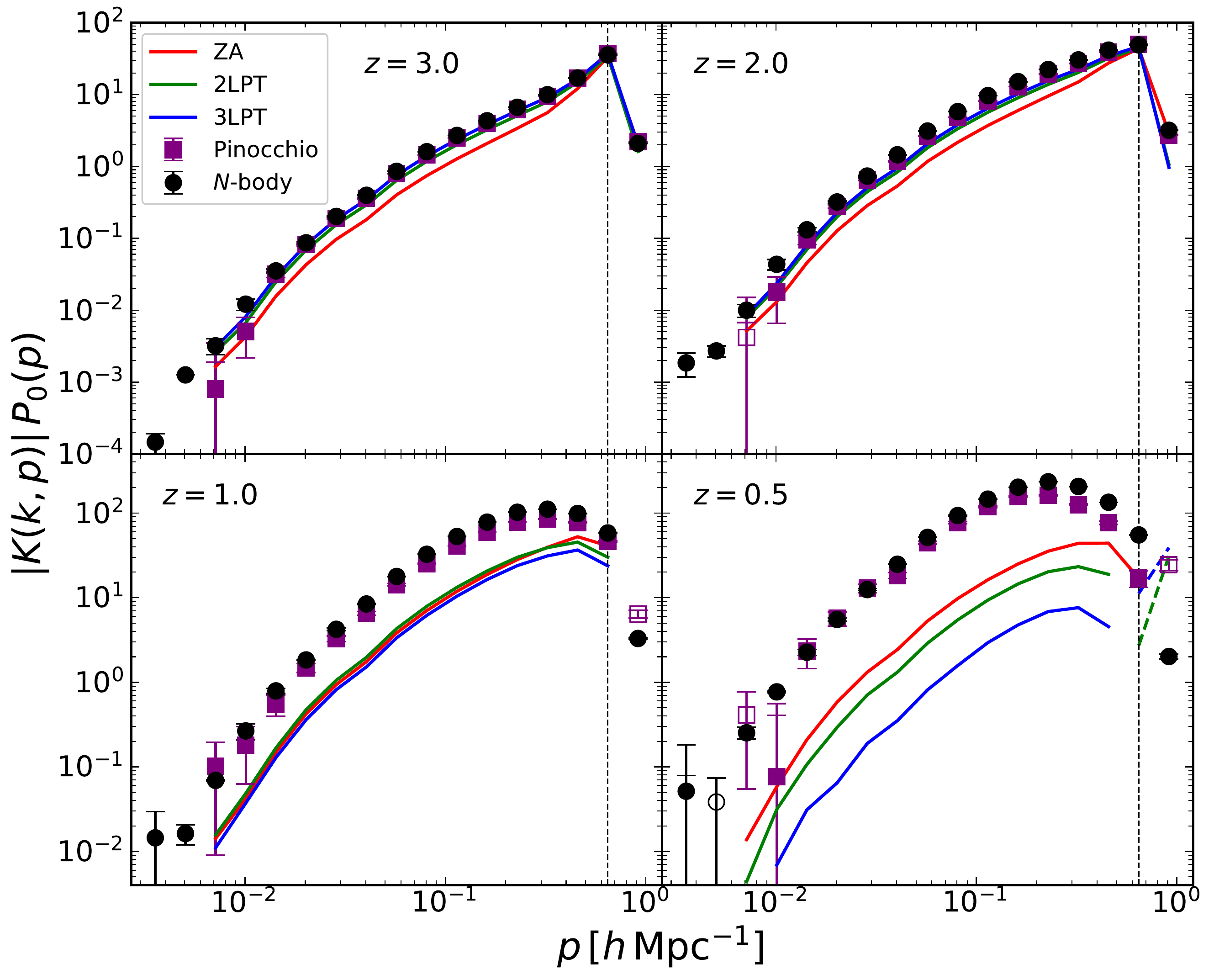}
\caption{Same as Fig.~\ref{3DResponse}, but at higher $k$ wavenumbers (from left to right and top to bottom, $k=0.228$, $0.323$, $0.457$ and $0.646 h$ Mpc${}^{-1}$).}
\label{3DResponse_highk}
\end{figure*}

Figures~\ref{3DResponse} and \ref{3DResponse_highk} show the response as a function of $p$ for different values of $k$ (marked by the location of the peak) at the four redshifts we consider. Note that the plotted results are the absolute values of the response function multiplied by $P_0(p)$, i.e., $|K(k,p)|P_0(p)$, which slightly differ from those shown in Figures~ \ref{1D-fig} and \ref{1Dbis-fig}. The line types and the symbols are the same as in Fig.~\ref{3Dpower}, except that we employ filled (open) symbols to show a positive (negative) value of $K(k,p)$. We use the default setting of $512^3$ mass elements for the \texttt{PINOCCHIO} realizations.

The behaviour of LPT predictions is somewhat different in the left part ($p<k$) and the right part ($p>k$) of each panel of the figures. On the left, LPT results get closer to $N$-body results when increasing the PT order, except when probing too deeply the non-linear regime (high values of $k$ and low $z$ in Fig.~\ref{3DResponse_highk}). However, the trend is not monotonic on the right side of the panels. While 2LPT is always worse than ZA in this region, 3LPT approaches $N$-body results at high redshifts. This is different at low redshifts, where 3LPT prediction performs the most poorly, especially for the highest $p$ bins, regardless of the value of $k$ in this range. This $k$-independent suppression of the mode coupling to UV is fully consistent with the finding of \citet{Nishimichi:2014rra}, despite the fact that their calculation is based on the Eulerian PT. The underperformance of 3LPT in predicting the matter power spectrum may be ascribed to a too strong sensitivity to ultraviolet (UV) contributions.

The \texttt{PINOCCHIO} prediction shown by the square symbols is the closest to the $N$-body data. It preserves the accuracy of 3LPT for $p<k$ and at the same time suppresses the strong UV sensitivity seen in 3LPT. On the smallest scales, however, there is still a sizeable mismatch between \texttt{PINOCCHIO} and $N$-body, which appears again to be mainly sensitive to $p$ and not to $k$.

\begin{figure*}
\centering
\resizebox{\hsize}{!}{
\begin{tabular}{cc}
\includegraphics{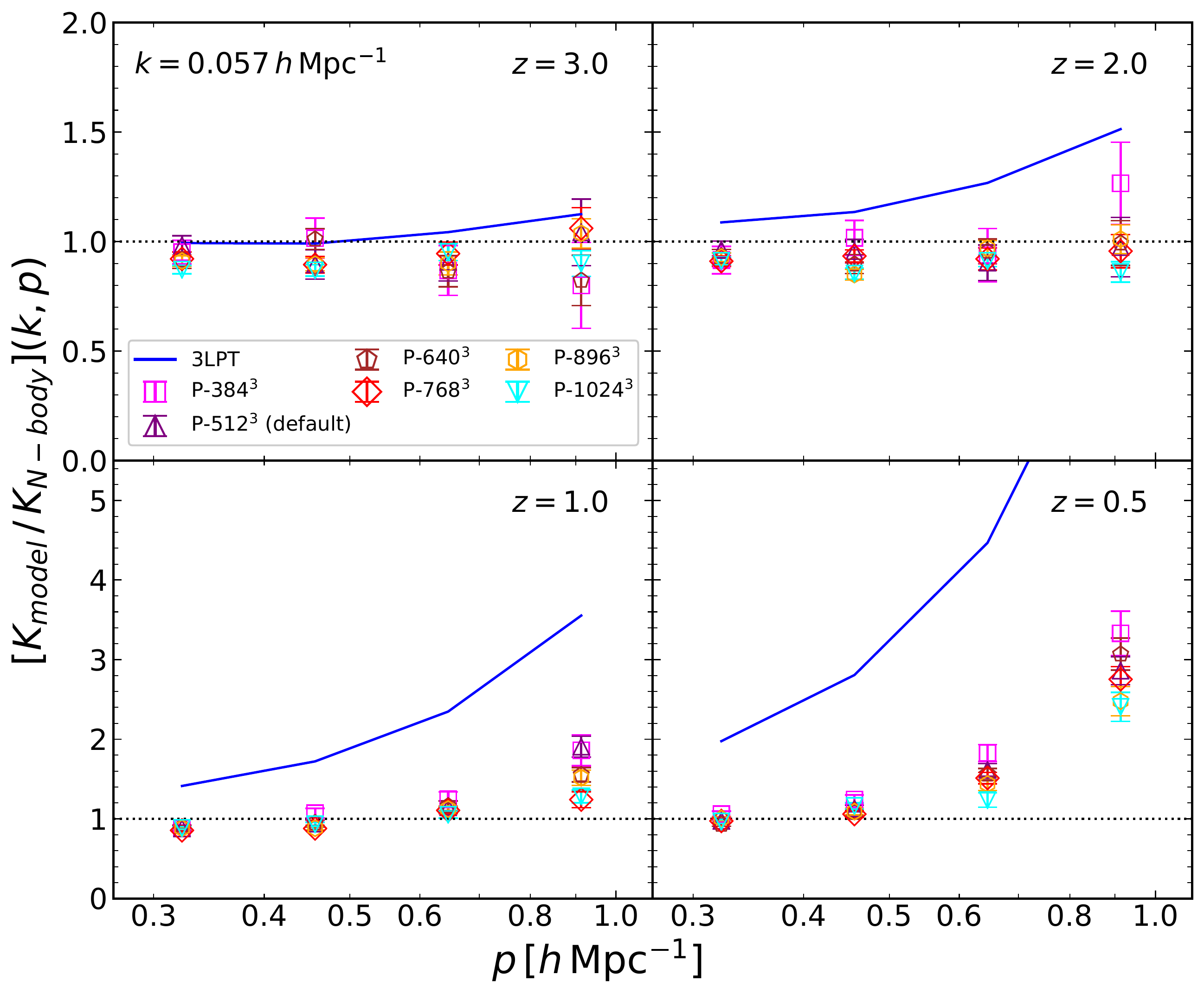} & \includegraphics{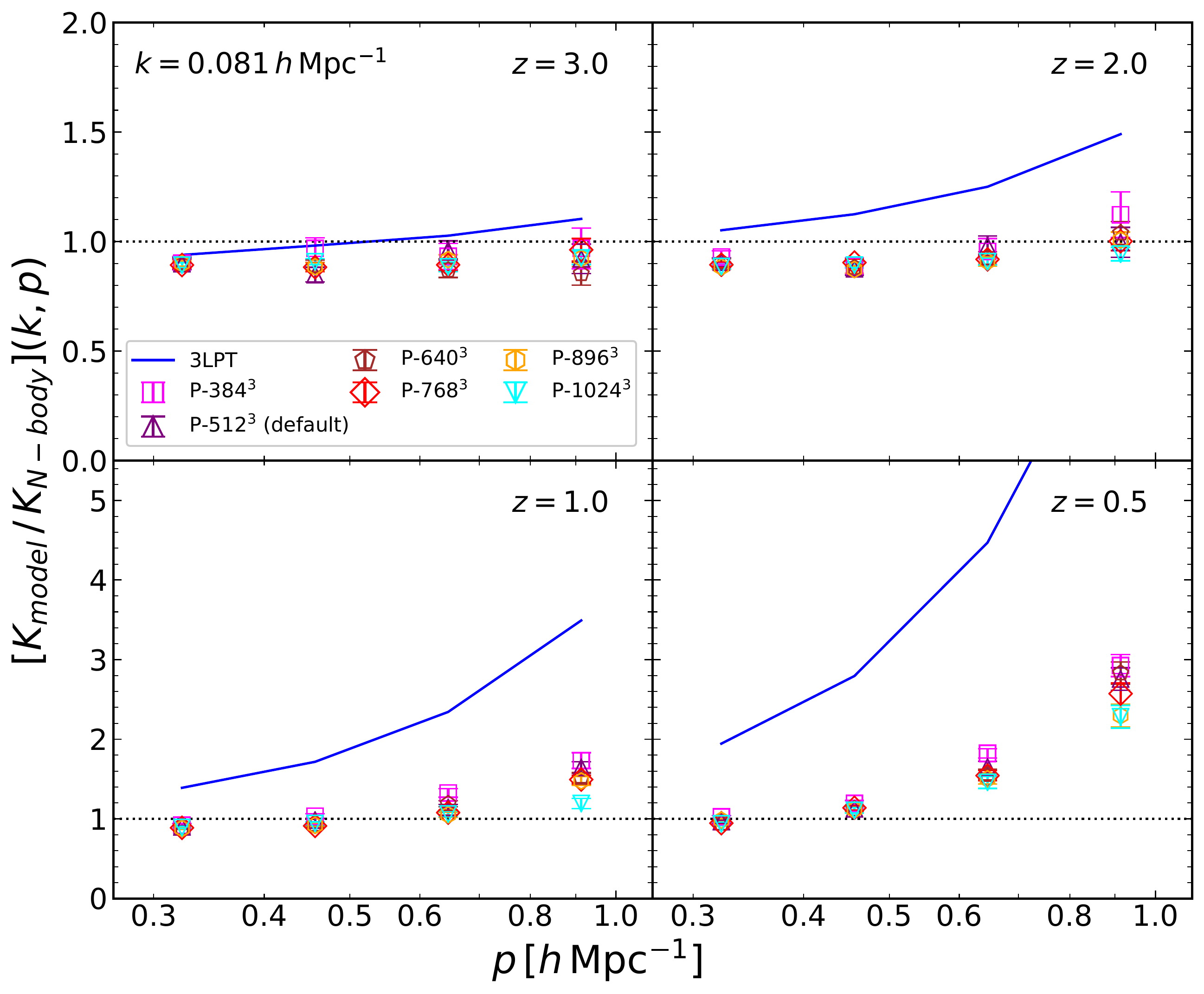} \\
\includegraphics{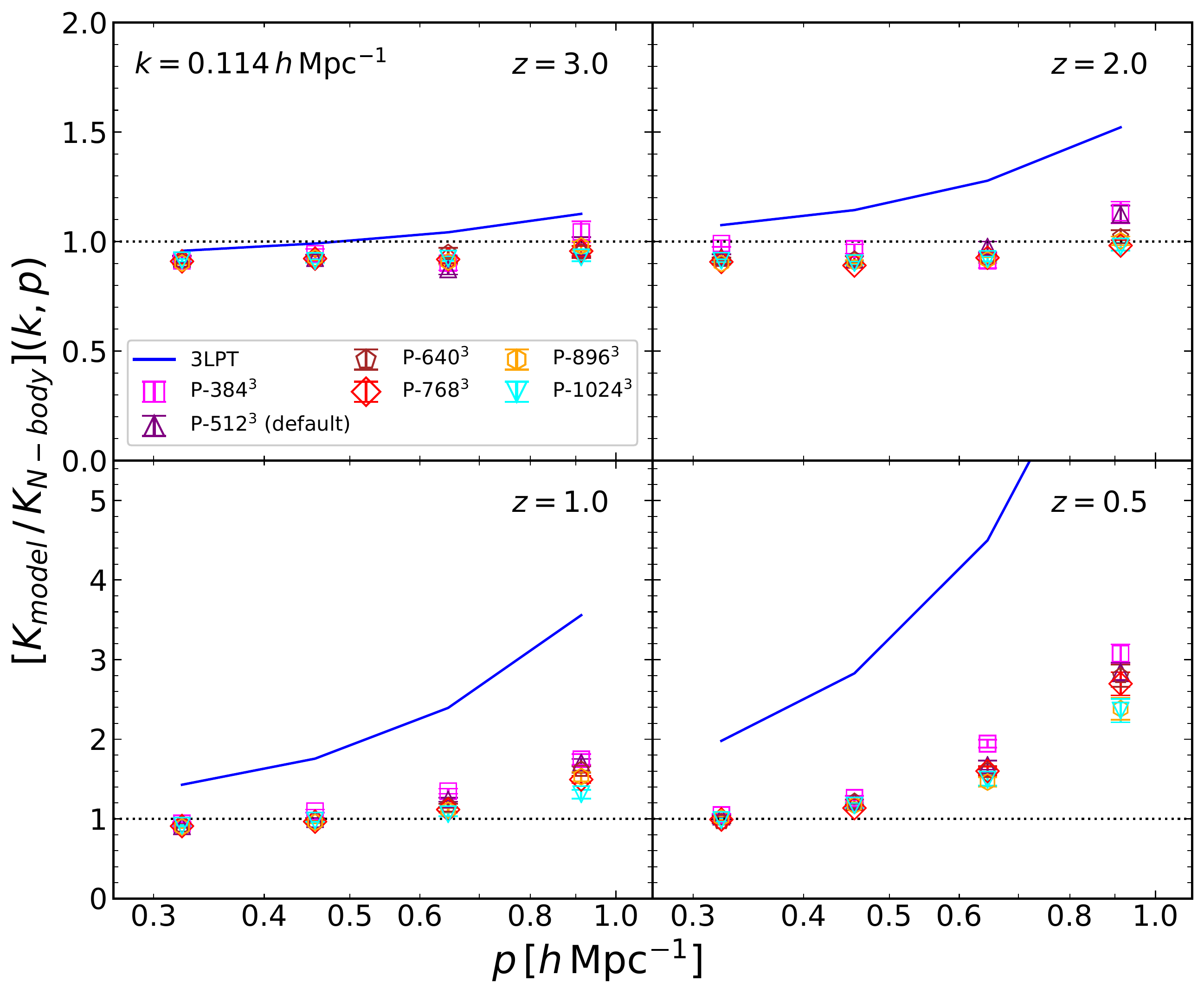} & \includegraphics{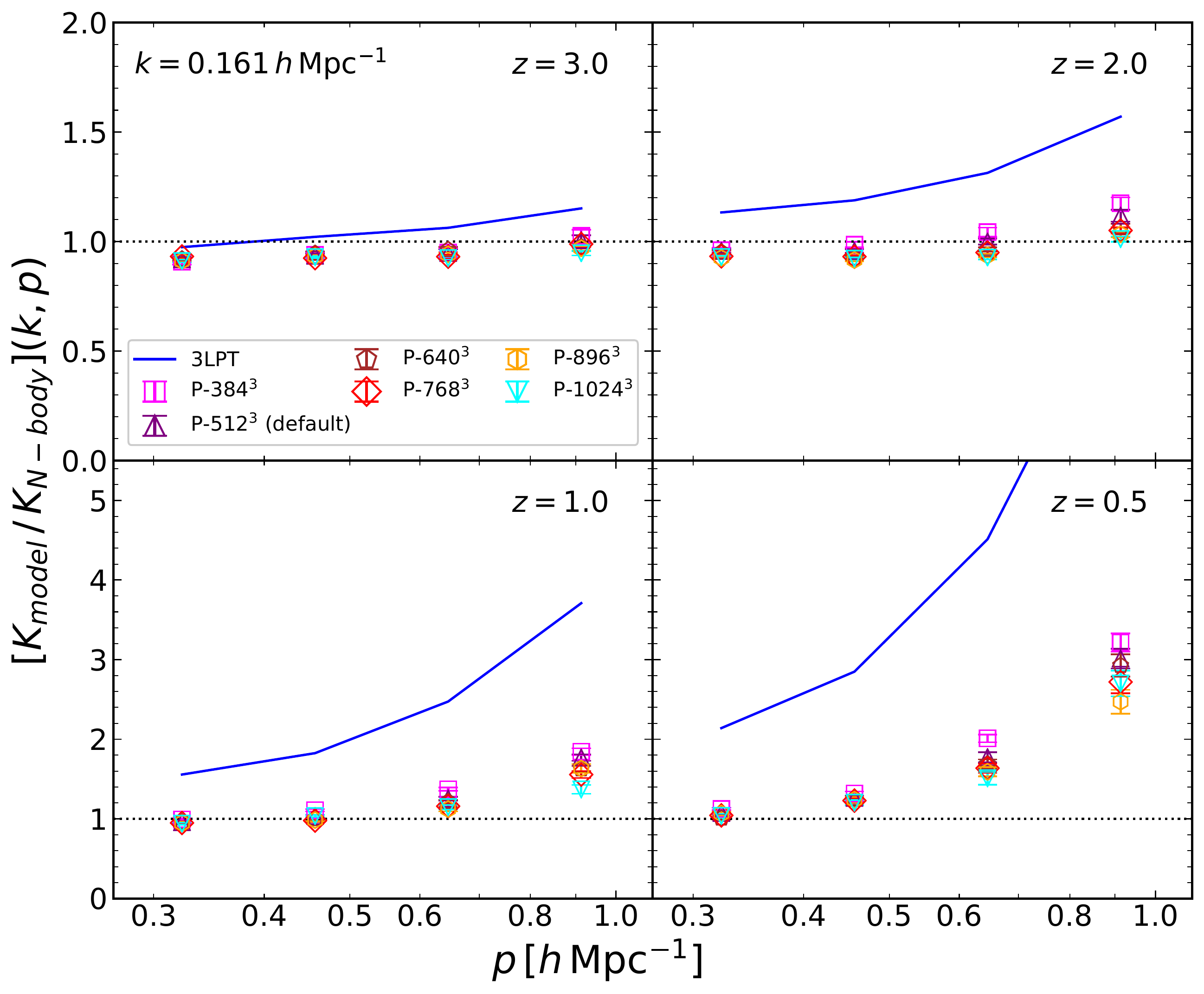}
\end{tabular}}
\caption{Convergence study of the \texttt{PINOCCHIO} prediction for the response function in the large-$p$ region of Fig.~\ref{3DResponse}. Results are normalized by the measurements in $N$-body simulations. Different symbols correspond to different mass resolutions of \texttt{PINOCCHIO}. As a reference, we also show with a solid line the result of 3LPT, which is also underlying to the \texttt{PINOCCHIO} realizations.}
\label{3DResponse_zoom}
\end{figure*}

Since \texttt{PINOCCHIO} can correct for the dynamics in haloes only above the mass resolution limit, its convergence must be carefully checked, particularly in the UV regime. Fig.~\ref{3DResponse_zoom} shows, for the $4$ largest $p$ bins in each panel of Fig.~\ref{3DResponse}, the ratio between the response function from \texttt{PINOCCHIO} with different mass resolutions (as detailed in legend of the figure) and the one from the $N$-body simulations. For comparison, we also show the ratio between pure 3LPT and $N$-body, since \texttt{PINOCCHIO} uses 3LPT dynamics for generating displacement of particles outside haloes.

While dependence on resolution is not always monotonic, probably in part because of the large error bars, a weak tendency that higher resolution realizations give a stronger suppression of the UV sensitivity can be observed. This is in agreement with intuition, since, by resolving smaller haloes, we can correct the multi-streaming dynamics of more mass elements. Although it is still not fully clear from this figure alone, the remaining visible discrepancies between the highest-resolution \texttt{PINOCCHIO} realizations and the $N$-body simulations suggest that correction of multi-streaming is incomplete. In particular, dynamics inside filaments and sheets is certainly not properly described by the treatment with mock haloes used in \texttt{PINOCCHIO}. Indeed, there is no known simple procedure to account for post-collapse dynamics in filaments and sheets and this can as well affect the way haloes are identified in \texttt{PINOCCHIO} through the merging tree procedure used to group multi-stream fluid elements. 

\begin{figure*}
\centering
\resizebox{\hsize}{!}{
\begin{tabular}{cc}
\includegraphics{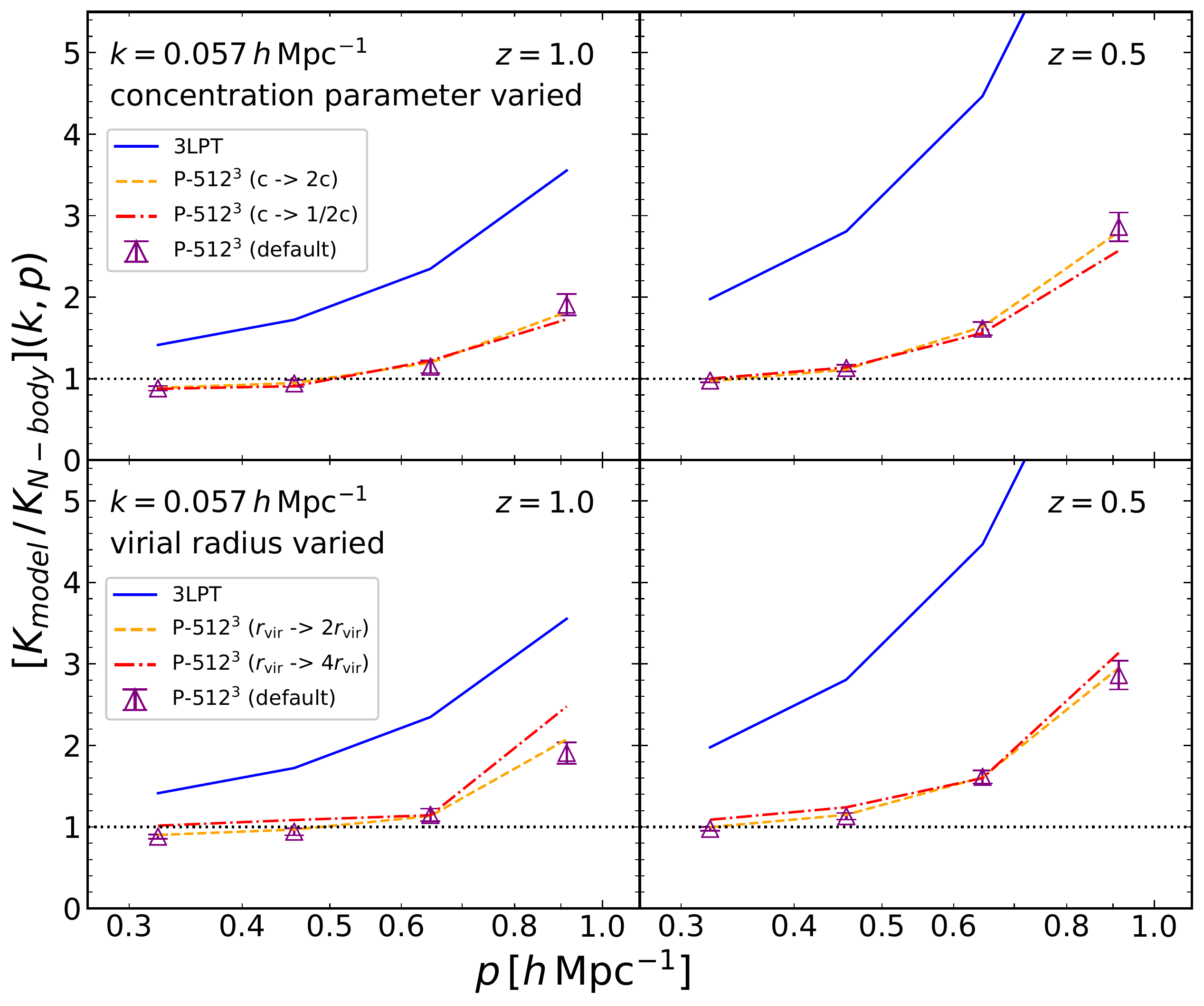} & \includegraphics{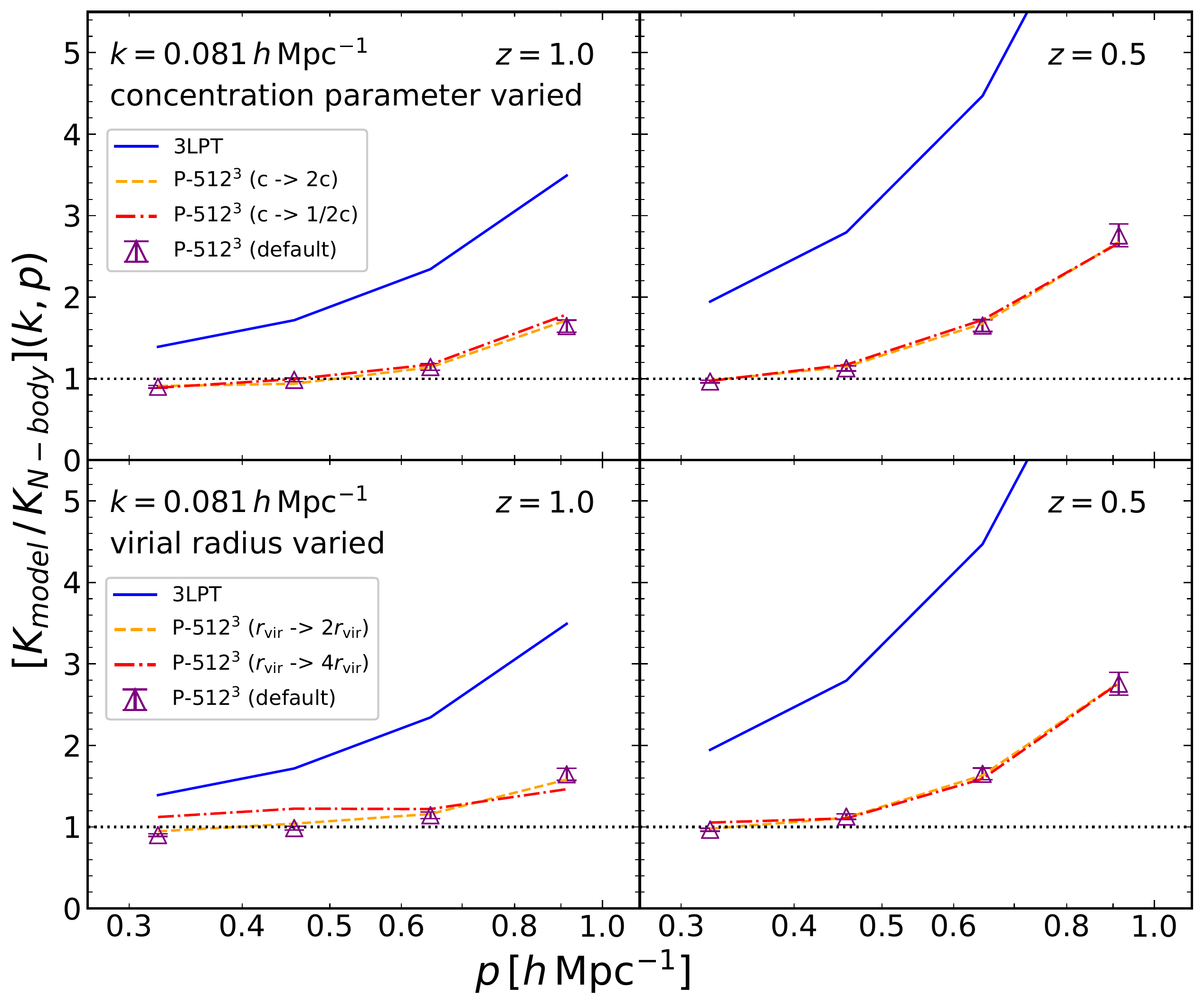} \\
\includegraphics{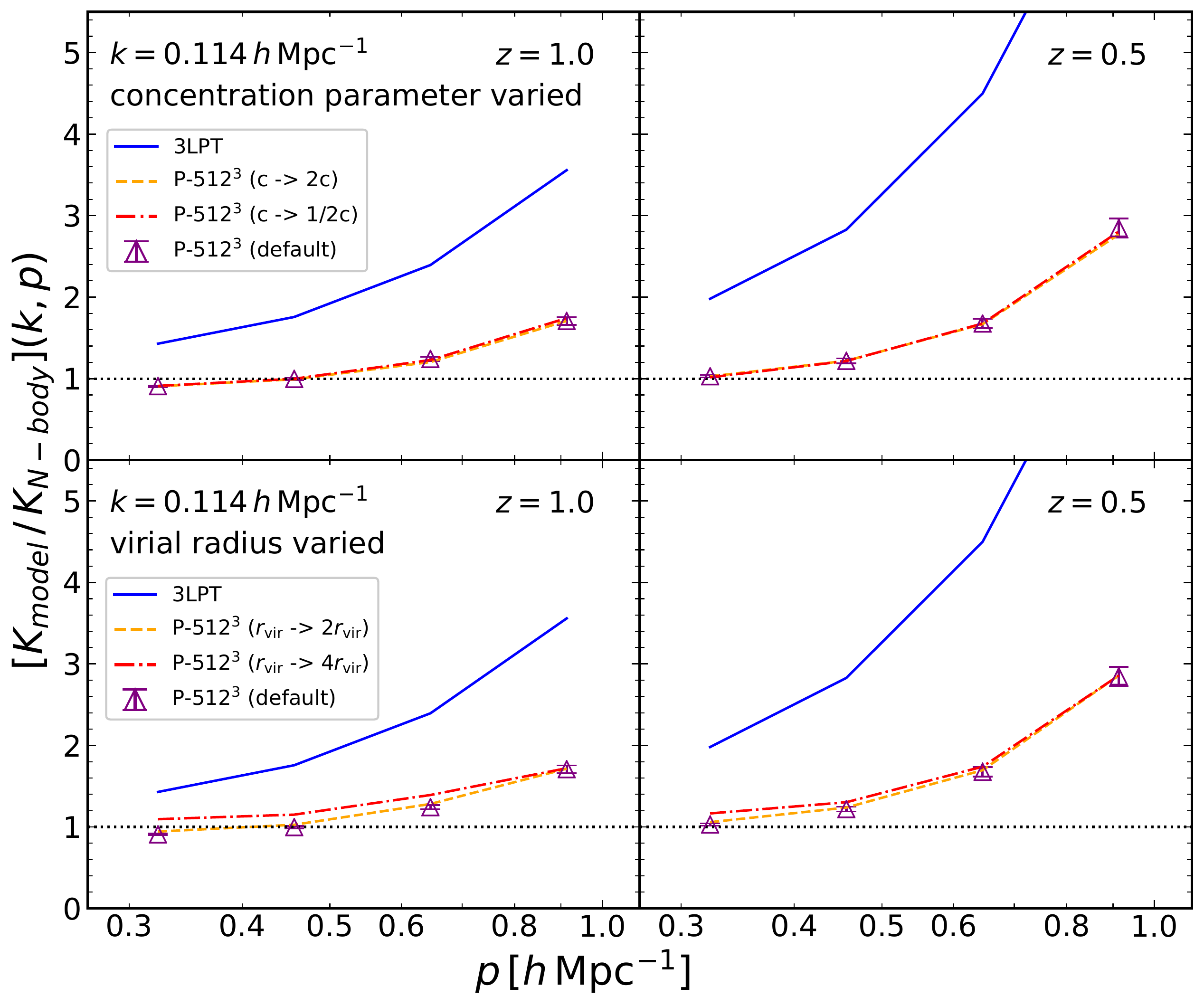} & \includegraphics{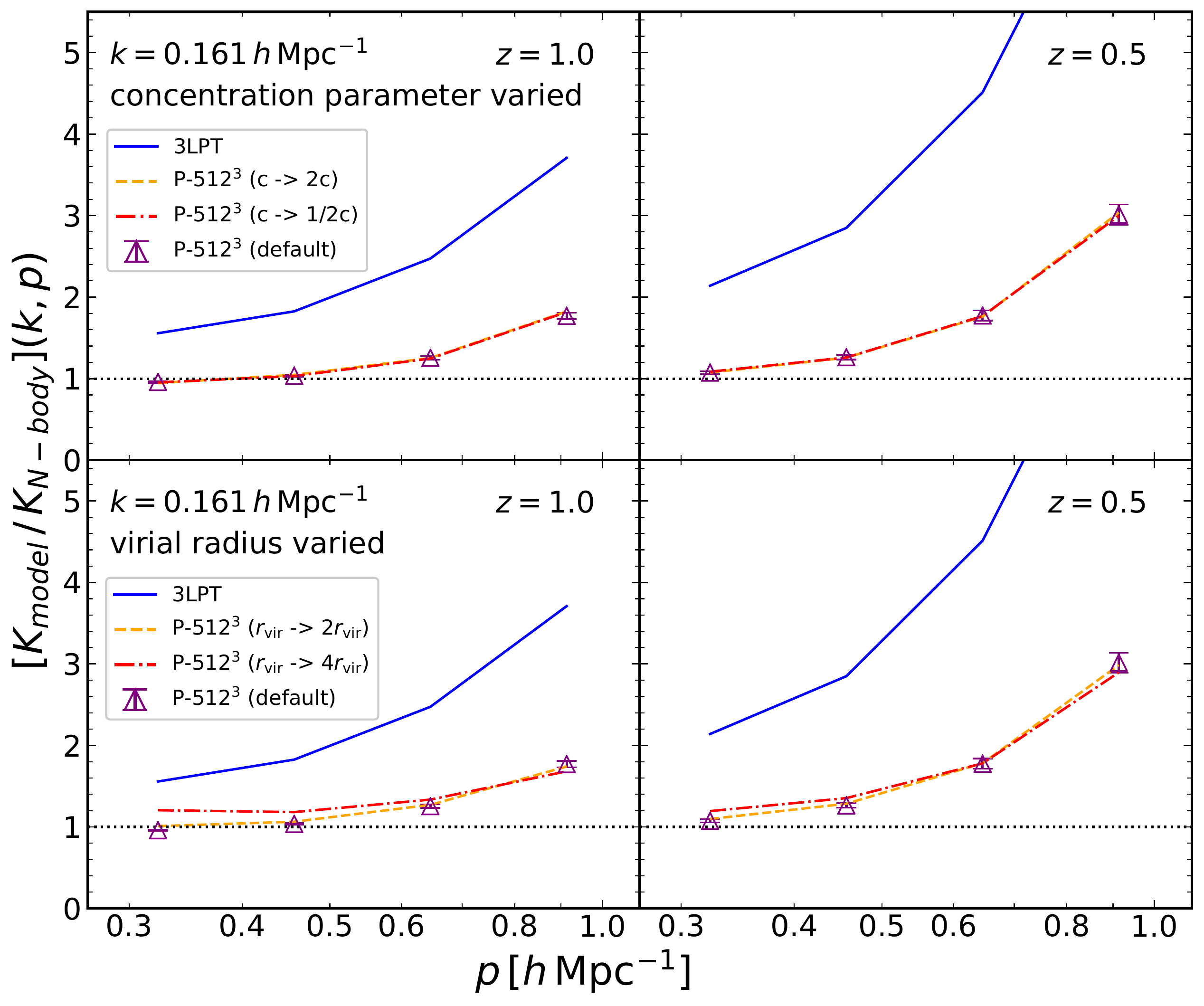}
\end{tabular}}
\caption{Similar to Fig.~\ref{3DResponse_zoom}, but for \texttt{PINOCCHIO} runs with modified halo profiles. We focus on $z=1$ and $0.5$, where non-perturbative effects are prominent, and show the results when the concentration parameter is modified (upper row in each panel) or the virial radius is modified (lower row in each panel) at four different wavenumbers $k$ as indicated in the figure legend. The results are compared to the default setting of \texttt{PINOCCHIO} (triangles with error bars) as well as 3LPT (solid lines). The results with modified halo parameters shown by the dashed or dot-dashed lines are almost on top of the triangles, suggesting that detailed mass profiles within haloes do not significantly change the response function in the quasi non-linear scales.}

\label{3DResponse_zoom_concentration}
\end{figure*}

Another possible source of discrepancy consists in replacing the ``halo'' particles with NFW spheres, which might seem to be rather crude. Indeed, it is natural to expect that in reality, a slight change in the initial linear power spectrum can lead to a change in the mass profile of haloes, giving additional contributions to the response function. To see this more in detail, we additionally perform \texttt{PINOCCHIO} runs but with the parameters describing the halo profile artificially modified, while still assuming sphericity. Firstly, we double (respectively halve) the concentration parameter by hand from the default \texttt{PINOCCHIO} implementation, employing the fitting formula by \citet{Bhattacharya_2013}. Secondly, we consider changing the value of the virial radius, $r_\mathrm{vir}$, to twice or four times the default value. In practice, we multiply the distance of the member particles from the center by two or four, and the resultant concentration parameter (defined as the ratio of the virial radius to the scale radius) is thus unchanged, while the over-density inside the ``haloes'' becomes significantly smaller than the standard value of $\sim200$. We prepare 100 pairs of simulations with modifications in the linear power spectrum at each of the four wavenumbers on non-linear scales, for each of the four non-standard settings described above. We show the results in Fig.~\ref{3DResponse_zoom_concentration} together with the 3LPT and the default \texttt{PINOCCHIO} runs. Despite the rather exaggerated change in the concentration parameter or the virial radius compared to the current numerical calibration level, it is clear from the figure that the response function is not affected beyond the error bars. This exercise indicates that the details of mass distribution inside haloes do not play a role in shaping the response function at quasi non-linear scales. Additionally, although we still assumed a spherical shape for the haloes, the extreme nature of this exercise suggests as well that accounting for halo ellipticity should not significantly influence the results at quasi non-linear scales, except maybe if there are cumulative effects, for example related to alignments of the haloes shapes with the cosmic web.

\begin{figure}
    \centering
    \includegraphics[width=8.5cm]{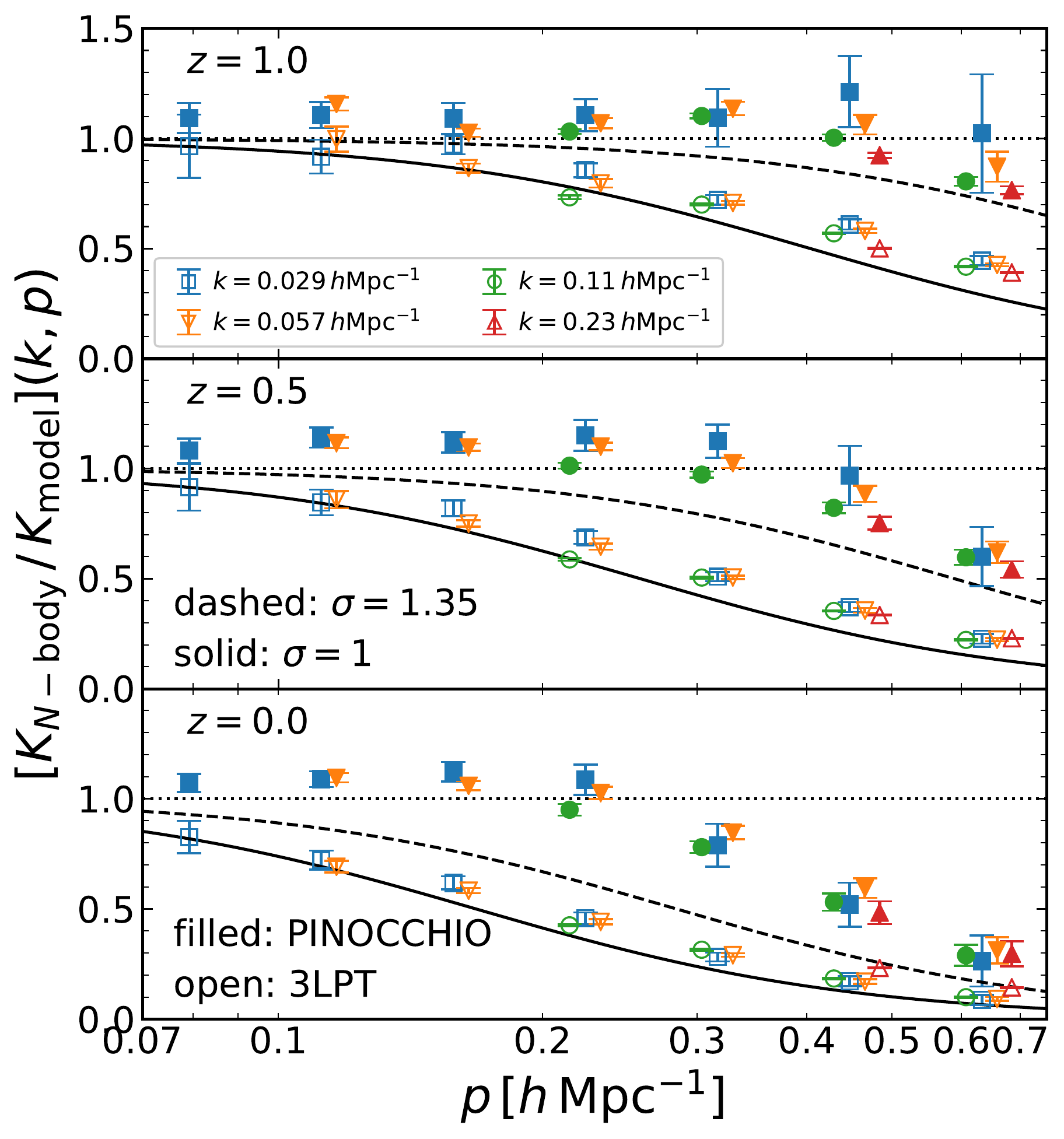}
    \caption{Damping of the response function seen in the $N$-body dynamics compared to  models. We show the ratio as a function of the wavenumber $p$ at various wavenumbers $k$ depicted by different symbols as indicated in the figure legend at three redshifts, $z=1$, $0.5$ and $0$ from top to bottom. We consider 3LPT (open) and \texttt{PINOCCHIO} (filled) as the model used in the denominator. Also shown are the fitting form proposed by \citet{Nishimichi:2014rra} (dashed line) and its variant with the $\sigma$ parameter changed to unity (solid). We only plot data points at $p\geq 2k$ to focus on the mode transfer from a smaller to a larger scale and different symbols are slightly shifted horizontally to avoid heavy overlap.}
    \label{fig:UV_damping}
\end{figure}

Finally, to make a closer connection to the UV sensitivity of the non-linear matter power spectrum previously reported in \cite{Nishimichi:2014rra}, we show the damped response seen in the full $N$-body simulations compared to the perturbative and non-perturbative models in Fig.~\ref{fig:UV_damping}. We plot the ratio of the response function measured from the $N$-body simulations to that from 3LPT by the open symbols. The figure indicates that the ratio decays towards larger $p$'s almost independently of $k$, which is indicated by different symbols. Furthermore, the decay looks more prominent at lower redshifts. The result should be compared to the dashed line, which shows the fitting form reported in \cite{Nishimichi:2014rra} based on the comparison between the $N$-body simulations and the two-loop SPT prediction. While the damping seen in the $N$-body/3LPT ratio is different in amplitude compared to the fitting form, the overall trend, especially the independence from $k$ and the trend with $p$ and $z$ are well recovered. Indeed, by adjusting the $\sigma$ parameter (defined in the introduction) in the fitting form, which characterizes the typical size of linearly extrapolated perturbations at which the damping takes place, we can explain the damping in the $N$-body/3LPT ratio. While the dashed line corresponds to $\sigma=1.35$ found in \cite{Nishimichi:2014rra}, we achieve a closer match by setting $\sigma=1$ (solid), which simultaneously explains the results at all the redshifts shown here\footnote{It would be possible to further fine-tune the value of $\sigma$ to better reproduce the numerical results. However, the precise value of $\sigma$ is not important to support our conclusions as long as it is around unity, and thus we do not pursue this any further.}. This suggests that the suppressed UV sensitivity in the fully non-linear dynamics compared to high-order PT-based predictions is rather universal despite the difference between the Eulerian and Lagrangian PT and/or the different order at which the PT series is truncated.

As already confirmed in previous plots, the better agreement between the model with non-perturbative corrections and the $N$-body results can be checked again in Fig.~\ref{fig:UV_damping}. The filled symbols show the same ratio but with \texttt{PINOCCHIO} in place of 3LPT. After applying the correction of particle trajectories beyond shell-crossing with this algorithm, the damping of the ratio is not seen at $z=1$ and only visible at large wavenumbers at lower redshifts ($p\gtrsim0.5\,h\mathrm{Mpc}^{-1}$ at $z=0.5$ and $p\gtrsim0.3\,h\mathrm{Mpc}^{-1}$ at $z=0$). This suggests that the crude modelling of particle trajectories in halo regions is sufficient to recover the response function, including its redshift and wavenumber dependence, up to certain values of $p$ depending on the redshifts, and that non-linear structures corresponding to the scales where the residual decay is still visible, especially at lower redshifts, are not perfectly regularized by this simple recipe. The residual damping in the $N$-body/\texttt{PINOCCHIO} ratio would give a rough idea of the mass scale of objects for which the effect of shell-crossing is not fully accounted for by the current model. Nevertheless, the current model provides a reasonable match to the response function over the wavenumbers covering the BAO scale until $z=0$, despite the rather crude way employed for the post-shell-crossing dynamics (i.e., relocation of particles to form NFW spheres) and we postpone further investigation of the regularization of the PT dynamics on even smaller scales.

We conclude from the tests in this section that the breakdown of LPT is strongly associated with the dynamics of mass elements after shell-crossing in three dimensions. This is explicitly shown in terms of the response function, in particular the sensitivity of the perturbations in the weakly non-linear regime to those at smaller, non-linear scales. A proper account of shell-crossed regions is a key to regularize LPT dynamics. Since haloes are the main sites where shell-crossing occurs, a large part of the UV sensitivity is alleviated by relocating the member particles to form a NFW sphere. In addition, we show that the detail of regularization is not crucial to obtain a well-behaved response function, as long as the particle trajectories after shell-crossing are confined in a reasonably small region.

\section{Conclusion}
\label{sec:conclusion}

In this paper, we have studied the power spectrum response function, $K(k,p)$, originally introduced in the 3D case by \citet{Nishimichi:2014rra} and \citet{Nishimichi_etal2017}. Analyses by these authors have found that the response function becomes negative at $k<p$, and that the absolute value of its amplitude in $N$-body simulations is suppressed compared to the prediction of perturbation theory (PT) if the mode $p$ enters the non-linear regime. That is, the actual mode coupling between small and large scales is suppressed, as opposed to the PT predictions. One can postulate that this suppressed mode coupling is due to the fact that multi-stream dynamics is not correctly accounted for by PT which is valid only until shell-crossing. In order to check this hypothesis, we compared measurements in $N$-body simulations to Lagrangian PT predictions augmented (or not) with approximate recipes accounting for multi-stream dynamics in the strongly non-linear regime. We first started with the idealistic case of 1D cosmology, in which the Zel'dovich solution is exact in the single-stream regime. In addition to the Zel'dovich approximation, we tested post-collapse PT developed by \citet{Colombi:2014lda} and \citet{Taruya_Colombi2017}, which is able to describe in an approximate way the local evolution from first to next crossing time. To account for highly non-linear evolution, including merger events, we supplemented Zel'dovich and post-collapse PT with an adaptive smoothing procedure of initial conditions to select haloes and summarize them as a ``S'' shape in phase-space with the proper size. Then we turned to the 3D case, where Lagrangian PT, which, at variance with Zel'dovich solution in 1D, can only approach the exact solution in an approximated way even prior to shell-crossing, was tested against $N$-body simulations. To account for multi-stream dynamics, we used the software \texttt{PINOCCHIO}, which supplements third order Lagrangian PT with an adaptive smoothing procedure to select haloes and represents them with a spherical universal NFW profile. The main results of our investigations can be summarized as follows:
\begin{enumerate}
 \item[(i)] Response functions measured in 1D $N$-body simulations present a structure quite similar to the 3D case: at high redshift, the function $K(k,p)$ exhibits a sharp peak around $k=p$; looking at its behaviour as a function of $p$ for a fixed $k$, its sign eventually flips at a certain wavenumber $p > k$, with an amplitude strongly suppressed compared to the peak; at low redshift, the sharp peak structure tends to disappear and the response function acquires a flatter shape.
 \item[(ii)] The theoretical 1D response function computed from the exact single-stream treatment (i.e., Zel'dovich solution) presents stronger mode coupling between small- and large-scale modes than $N$-body simulations, similarly to what can be observed in the 3D case with Eulerian PT \citep{Nishimichi:2014rra, Nishimichi_etal2017} or our Lagrangian PT measurements, as illustrated by Fig.~\ref{3DResponse}.
 \item[(iii)] In the one dimensional case, post-collapse PT, which accounts partly for multi-stream dynamics, only improves slightly on Zel'dovich solution. On the other hand, when supplemented with the adaptive smoothing procedure, a substantial improvement of the agreement between post-collapse PT and $N$-body simulations results is found, even in the highly non-linear regime, $k,p\sim1 $\,Mpc$^{-1}$. Adaptive smoothing also improves Zel'dovich solution but not quite as well in the large $k$ and $p$ regime.
 \item[(iv)] In the three-dimensional case, LPT performs increasingly better with order in the regime $p < k$, as expected, but, as mentioned in (ii), overestimates mode couplings for $p > k$. Third-order LPT in fact behaves worse than second-order LPT and Zel'dovich approximation at low redshifts and for large $p$. The improvements brought by \texttt{PINOCCHIO} over pure LPT are significant in the regime $p > k$. The results are robust against significant changes in the parameters describing the mass profiles of the haloes in \texttt{PINOCCHIO}. However, the amount of mode coupling suppression for $p > k$, after a convergence study, is found to be still insufficient to quantitatively match $N$-body results. This can be explained, at least in part, by the fact that multi-stream dynamics inside non-linear structures such as filaments and sheets is certainly not described accurately enough with the merging tree procedure implemented in \texttt{PINOCCHIO}. We discuss this further below.
\end{enumerate}
These findings readily imply that the suppressed mode coupling between small- and large-scale structure is intimately related to the dynamics of small-scale clustering after shell-crossing, and a proper way to describe the multi-stream flows is important to account for the large-$p$ behaviour of the response function. Indeed, because single stream PT does not account for counter terms on the force field inside multi-stream regions, it introduces artificial couplings between small- and large-scale modes. However, the results above show that it is not necessary to follow all the details of highly non-linear dynamics to correct for these defects, but rather to provide a reasonable modelling of multi-stream regions while preserving the bulk properties of the matter distribution outside them. In order to reach this objective, PT can be supplemented with an adaptive-smoothing procedure. The smoothing scale depends on a local crossing time, similarly to the excursion set approach.

In 1D, combining post-collapse PT with such an adaptive smoothing technique is shown to reproduce remarkably well the measured response function measured in $N$-body simulations. The 3D case is far more complicated because there are locally three directions of motion, as illustrated, e.g. by Zel'dovich dynamics or ellipsoid collapse. The first structures to form are pancake like but their subsequent evolution is complex. The condition for the formation of a halo is actually far from trivial, although one could intuitively relate this event to shell-crossings occurring along all the three major axes of local motion. In \texttt{PINOCCHIO}, multi-stream regions are identified and drawn in Lagrangian space by using the first crossing time found by combining ellipsoid collapse dynamics with adaptive, isotropic smoothing. Already, one might question isotropic smoothing since there are preferred directions of local motion, although these latter are precisely taken care of by ellipsoid dynamics. Then, haloes inside these regions are identified with some friend-of-friend procedure combined with second order LPT for tracing motion of element of fluids. This process basically describes the dynamics along the two directions of motion not treated yet, necessary to really define a halo from the dynamical point of view. This procedure remains very approximate since it does not give account of counter terms in the force field inside the multi-stream regions, but is calibrated with $N$-body simulations to have the best possible matching of the halo mass function. However, even if haloes are correctly identified thanks to the calibration step, their position remains approximate because internal dynamics of filaments and sheets is in fact not accounted for accurately enough. Improvement of this step might represent one of the key points to remedy the mismatch observed between \texttt{PINOCCHIO} and the $N$-body simulations for the response function at large $p$. Post-collapse PT, if applicable to 3D, should be able, as we have seen in 1D when comparing it to Zel'dovich dynamics, to provide at least partial answers to this issue. Finally, while our analyses suggest that the results do not significantly depend on the details of the supposed halo shape, it is possible that accounting for non sphericity of the haloes and how they align with the structures that host them, such as clusters, filaments and sheets might furthermore improve the results.

\section*{Acknowledgements}
This work was supported in part by ANR grant ANR-13-MONU-0003 (AH \& SC) and by MEXT/JSPS KAKENHI grants JP15H05889, JP16H03977 (AT), JP17K14273 and JP19H00677 (TN), Japan Science and Technology Agency (JST) CREST JPMHCR1414 (TN) and JST AIP Acceleration Research Grant Number JP20317829, Japan (TN \& AT). Numerical computation was carried out in part using the HPC resources of CINES (Occigen supercomputer) under the GENCI allocation 2018-A0040407568, and on Cray XC50 at Center for Computational Astrophysics, National Astronomical Observatory of Japan. It has also made use of the Yukawa Institute Computer Facility.

%

\appendix
\section{Power spectrum and response function from Zel'dovich solution}
\label{app:predictions_zeldovich}

In this Appendix, we derive analytical expressions for the power spectrum (Sec.~\ref{subsec:power_spectrum_ZA}) and the response function (Sec.~\ref{subsec:response_func_ZA}) based on the 1D Zel'dovich solution.

\subsection{Power spectrum}
\label{subsec:power_spectrum_ZA}

The analytic expression of the 1D power spectrum for Zel'dovich solution has been given in the literature \citep[e.g.,][]{BondCouchman1986,1995MNRAS.273..475S,McQuinn:2015tva}. Here, for later convenience, we provide the detailed derivation in a self-contained manner.

Let us write down the Zel'dovich solution in 1D cosmology. Denoting the Eulerian position of mass elements by $x$, the Zel'dovich solution relates $x$ to (initial) Lagrangian position $q$ through
\begin{align}
x(q;z)=q+D_+(z)\psi(q),
\label{eq:Zel_sol}
\end{align}
where $\psi$ is the displacement field. The density field in Eulerian space, $\delta(x)$, is related to the mass density field in Lagrangian space through Eq.~(\ref{eq:Zel_sol}). Since the Lagrangian mass density field is supposed to be homogeneous, we have
\begin{align}
\delta(x)&=\left|\frac{\partial x}{\partial q}\right|^{-1}-1
\nonumber\\
&=\frac{1}{1+D_+\,({\rm d}\psi/{\rm d}q)} -1.
\label{eq:delta_x}
\end{align}
At early time, $D_+\ll1$, the above equations indicate $x\simeq q$
and $\delta(x)\simeq-D_+({\rm d}\psi/{\rm d}q)$. Denoting the
Eulerian linear density field by $\delta_0$ (proportional to initial density field), we have
\begin{align}
\delta_0(x=q)=-D_+\,\frac{{\rm d}\psi}{{\rm d}q}.
\label{eq:delta_0}
\end{align}

We are interested in Fourier space statistics.
To be specific, consider the power spectrum defined by
\begin{align}
\bigl\langle\delta(k)\delta(k')\bigr\rangle=2\pi\,\delta_{\rm D}(k+k')\,P^{\rm (1D)}(k).
\label{eq:def_pk}
\end{align}
To derive the expression for the power spectrum based on the Zel'dovich solution,
we first take the Fourier transform of Eq.~(\ref{eq:delta_x}):
\begin{align}
\delta(k)&=\int_{-\infty}^{\infty} {\rm d}x \,e^{ikx} \delta(x)
\nonumber\\
&=\int_{-\infty}^{\infty} {\rm d}q \,e^{ik\{q+D_+\psi(q)\}}-2\pi\,\deltad(k).
\label{eq:delta_k}
\end{align}
Note that for $D_+\ll1$, this reduces to
\begin{align}
\delta_0(k)&=ik D_+\int_{-\infty}^{\infty} {\rm d}q \,e^{ikq} \,\psi(q)
\nonumber\\
&\Longleftrightarrow D_+\,\psi(q)=\int_{-\infty}^{\infty}\frac{{\rm d}k}{2\pi}\,
e^{-ikq}\left\{-\frac{i}{k}\,\delta_0(k)\right\}.
\label{eq:delta_0_k}
\end{align}
This is consistent with the expression in Eq.~(\ref{eq:delta_0}).
Now, with the density field given above,
let us evaluate the left-hand side of Eq.~(\ref{eq:def_pk}). We have
\begin{align}
\bigl\langle\delta(k)\delta(k')\bigr\rangle &=\int {\rm d}q\int {\rm d}q'
\Biggl\langle
\Bigl\{e^{ik\{q+D_+\psi(q)\}}-e^{ikq}\Bigr\}
\nonumber
\\
&\qquad \times\Bigl\{e^{ik'\{q'+D_+\psi(q')\}}-e^{ik'q'}\Bigr\}
\Biggr\rangle
\nonumber\\
&= \int {\rm d}q\int {\rm d}q' e^{i(kq+k'q')}\Biggl[
\Bigl\langle e^{i\{k\psi(q)+k'\psi(q')\}D_+}\Bigr\rangle
\nonumber
\\
&\qquad
-\Bigl\langle e^{ik \,D_+\psi(q)}+e^{k'\,D_+\psi(q')}\Bigr\rangle+1\Biggr]
\nonumber\\
&= \int {\rm d}q\int {\rm d}q' e^{i(kq+k'q')} \nonumber
\\
&\qquad \Biggl[ \Bigl\langle e^{i\{k\psi(q)+k'\psi(q')\}D_+}\Bigr\rangle-1\Biggr].
\end{align}
Here, in the last line, we have used the fact that $\langle e^{ikD_+\psi(q)}\rangle$ becomes independent of $q$. To proceed further, we replace the integration
variables with $Q\equiv(q+q')/2$, $\Delta q\equiv q-q'$. We then obtain
\begin{align}
\bigl\langle\delta(k)\delta(k')\bigr\rangle
&=\int {\rm d}Q \int {\rm d}\Delta q \,e^{i(k+k')Q + i(k-k')\Delta q/2}
\nonumber
\\
&\qquad \times
\Biggl[\Bigl\langle e^{i\{k\psi(q)+k'\psi(q')\}D_+}\Bigr\rangle-1\Biggr]
\nonumber\\
&=(2\pi)\,\deltad(k+k')\,\int {\rm d}\Delta q\,
\,e^{ik\Delta q}
\nonumber
\\
&\qquad \times
\Biggl[\Bigl\langle e^{ik\{\psi(q)-\psi(q')\}D_+}\Bigr\rangle-1\Biggr].
\end{align}
Note that the integrand in the bracket depends only on $\Delta q$.
The comparison with Eq.~(\ref{eq:def_pk}) then leads to
\begin{align}
P^{\rm(1D)}_{\rm ZA}(k) &=\int {\rm d}\Delta q \,e^{i k\Delta q}
\Biggl[\Bigl\langle e^{ik\{\psi(q)-\psi(q')\}D_+}\Bigr\rangle-1\Biggr].
\label{eq:pk_ZA_general}
\end{align}
The above expression is still general in the sense that the statistical
correlations of the displacement field in the exponent are not yet specified.
In the following, we assume Gaussianity of linear density field,
$\delta_0$. In the Zel'dovich solution, this is equivalent to assuming
Gaussianity of the displacement field. Then, Eq.~(\ref{eq:pk_ZA_general})
becomes
\begin{align}
P^{\rm(1D)}_{\rm ZA}(k) &=\int {\rm d}\Delta q \,e^{i k\Delta q}
\Bigl[ e^{-k^2\langle\{\psi(q)-\psi(q')\}^2D_+^2\rangle/2}-1\Bigr].
\label{eq:pk_ZA_gauss}
\end{align}
Using Eq.~(\ref{eq:delta_0_k}), the displacement field correlation is
calculated to give
\begin{align}
\Bigl\langle\{\psi(q)-\psi(q')\}^2\Bigr\rangle D_+^2 &=
2\Bigl\{\bigl\langle\{\psi(q)\}^2 \bigr\rangle -\bigl\langle\psi(q) \psi(q')\bigr\rangle\Bigr\}D_+^2
\nonumber\\
&=
2\int\frac{{\rm d}k}{2\pi}\int\frac{{\rm d}k'}{2\pi}
\nonumber \\
&\qquad \Bigl\{e^{-(k+k')q}-\,e^{-(kq+k'q')}\Bigr\}\,
\nonumber
\\
&\qquad \times\frac{1}{kk'}\Bigl\langle\delta_0(k)\delta_0(k')\Bigr\rangle.
\end{align}
Defining the linear power spectrum $P_0$ by $\langle\delta_0(k)\delta_0(k')\rangle=2\pi\,\deltad(k+k')\,P_0(k)$, we get
\begin{align}
\Bigl\langle\{\psi(q)-\psi(q')\}^2\Bigr\rangle D_+^2 =
2\int_{-\infty}^\infty\frac{{\rm d}k}{2\pi}
\Bigl\{1-\,e^{-ik\Delta q)}\Bigr\}\,\frac{P_0(k)}{k^2}.
\end{align}
Substituting this into Eq.~(\ref{eq:pk_ZA_gauss}), we finally arrive at equations (\ref{eq:pk_ZA}) and (\ref{eq:func_Iq}) of Sec.~\ref{sec:zelres}.

\subsection{Response function}
\label{subsec:response_func_ZA}

Given the explicit functional form of the power spectrum in Eq.~(\ref{eq:pk_ZA}), we now derive the analytic expression for the response function $K^{\rm (1D)}_{\rm ZA}(k,p)$, Eq.~(\ref{eq:kernel_ZA_1Dmain}) in Sec.~\ref{sec:zelres}.

Consider a small variation of the linear power spectrum, $P_0\to P_0+\delta P_0$. Based on the Zel'dovich power spectrum in Eq.~(\ref{eq:pk_ZA}), the output response on the non-linear power spectrum reads
\begin{align}
\delta P^{\rm (1D)}_{\rm ZA}(k)&=\int_{-\infty}^\infty {\rm d}q\,e^{ikq}
(-k^2 ) \nonumber \\
& \qquad \Bigl\{\delta I(0)-\delta I(q)\Bigr\}e^{-k^2\,\{I(0)-I(q)\}}
\nonumber\\
&= \int_{-\infty}^\infty {\rm d}q\,e^{ikq} \nonumber \\
& \qquad
\Biggl(-k^2 \int_{0}^\infty \frac{{\rm d}p}{\pi} \,\Bigl\{1-\cos(pq)\Bigr\}\,
\frac{\delta P_0(p)}{p^2}\,\Biggr)
\nonumber\\
&\qquad\times
e^{-k^2\,\{I(0)-I(q)\}}
\nonumber\\
&=\int_0^\infty {\rm d}p\,
\Biggl(-\frac{2}{\pi}\frac{k^2}{p^2}\, \int_{0}^\infty {\rm d}q \,\cos(kq)\,\Bigl\{1-\cos(pq)\Bigr\}\,
\nonumber\\
&\qquad \times
e^{-k^2\,\{I(0)-I(q)\}}\Biggr)\,\delta P_0(p).
\label{eq:delta_P_ZA}
\end{align}
Remembering the definition of the response function of the power spectrum at wavenumber $k$ with respect to the initial small disturbance at wavenumber
$p$, $K_{\rm ZA}^{\rm(1D)}(k,p)$:
\begin{align}
\delta P^{\rm (1D)}_{\rm ZA}(k)=\int {\rm d}\ln p \,K_{\rm ZA}^{\rm (1D)}(k,\,p)\,\delta P_0(p),
\end{align}
Eq.~(\ref{eq:delta_P_ZA}) gives, using the property $I(-q)=I(q)$,
\begin{align}
K_{\rm ZA}^{\rm (1D)}(k,\,p)&=
-\frac{2}{\pi}\frac{k^2}{p}\, \int_{0}^\infty {\rm d}q \,\cos(kq)\,\Bigl\{1-\cos(pq)\Bigr\}\,
\nonumber\\
&\qquad \times e^{-k^2\,\{I(0)-I(q)\}}
\nonumber\\
&=-\frac{2}{\pi}\frac{k^2}{p}\, \int_{0}^\infty {\rm d}q \,\Bigl\{\cos(kq)\,-\frac{1}{2}\cos[(k-p)q]
\nonumber
\\
&\qquad -\frac{1}{2}\cos[(k+p)q]\Bigr\}\,e^{-k^2\,\{I(0)-I(q)\}}
\nonumber\\
&=-\frac{1}{\pi}\frac{k^2}{p}\, \int_{-\infty}^\infty {\rm d}q \nonumber \\
& \qquad \Bigl[e^{i\,kq}\,-\frac{1}{2}\left\{e^{i\,(k-p)q}+e^{i\,(k+p)q}\right\}\,\Bigr]
\nonumber
\\
&\qquad \times\,e^{-k^2\,\{I(0)-I(q)\}}.
\label{eq:K_ZA_1D}
\end{align}
Note that the response function should generally contain a contribution involving a Dirac delta function. In order to derive such a contribution, we utilize the fact that for Gaussian initial conditions, the power spectrum can in general be expressed in the following expansion form \citep{Bernardeau:2008fa,Bernardeau:2011dp,Taruya:2012ut}:
\begin{align}
 P^{\rm(1D)}(k) & = \{\Gamma^{(1)}(k)\}^2\,P_0(k)
\nonumber
\\
& + \sum_{n=2}^{\infty} \int \frac{{\rm d}p_1\cdots {\rm d}p_n}{(2pi)^{n-1}}\,\delta_{\rm D}(k-p_1-\cdots-p_n)
\nonumber
\\
&\times \{\Gamma^{(n)}(p_1,\cdots,p_n)\}^2\,P_0(p_1)\cdots P_0(p_n).
\label{eq:pk_Gamma_expansion}
\end{align}
where function $\Gamma^{(n)}$ is called the $(n+1)$-propagator, defined through the following functional derivative:
\begin{align}
\frac{1}{n!}\,\Bigl\langle\frac{\delta^n\,\delta(k)}{\delta\delta_0(p_1)\cdots\delta\delta_0(p_n)}\Bigr\rangle
&=\frac{1}{(2\pi)^{n-1}}\deltad(k-p_1\cdots-p_n)
\nonumber
\\
&\,\times\,\Gamma^{(n)}(p_1,\cdots,p_n).
\end{align}
From Eq.~(\ref{eq:pk_Gamma_expansion}), we can identify the contribution involving a Dirac delta function to the response function:
\begin{align}
K^{\rm (1D)}(k,\,p)=p\,\Bigl\{\Gamma^{(1)}(k)\Bigr\}^2\,\deltad(k-p) + \cdots,
\end{align}
where the two-point propagator $\Gamma^{(1)}$ can be analytically computed in the Zel'dovich solution, to give $\Gamma^{(1)}(k)=e^{-k^2I(0)/2}$. Adding this term to Eq.~(\ref{eq:K_ZA_1D}) and using Eq.~(\ref{eq:pk_ZA}), one obtains the full expression for the response function given in Eq.~(\ref{eq:kernel_ZA_1Dmain}) of Sec.~\ref{sec:zelres}.

Note that the integrals in the last two terms of Eq.~(\ref{eq:kernel_ZA_1Dmain}) still contain a Dirac $\delta$-function, $\deltad(k+p)$, which is out of the domain of definition of $K^{\rm(1D)}$. On the other hand, assuming the initial power spectrum $P_0(k)=k^2P_{\rm 3D}(k)/(2\pi)$ with $P_{\rm 3D}$ being the 3D matter power spectrum in the standard $\Lambda$CDM model, the function $I(0)-I(q)$ asymptotically behaves like $I(0)-I(q)\propto q^2$ at low-$q$, and approaches a constant in the high-$q$ limit. To be precise, $I(q)\to0$ in the high-$q$ limit, and $I(0)-I(q)$ approaches $I(0)$. This implies that the integrand in the third term of Eq.~(\ref{eq:kernel_ZA_1Dmain}) becomes vanishing in the high-$q$ limit, ensuring the convergence of the integral even in the vicinity of $k=p$.




\bibliographystyle{mnras}

%
\bibliography{ref}

\label{lastpage}
\end{document}